\documentclass[nofootinbib,preprint,preprintnumbers,a4paper,10pt]{revtex4}
\usepackage{multirow}
\usepackage{textcomp}
\usepackage{amsmath,graphicx,color,epsfig}

\usepackage{footnote}
\usepackage{ulem}
\usepackage{booktabs}
\usepackage{array}
\usepackage{amssymb}
\usepackage{subfigure}
\usepackage{longtable}
\usepackage{verbatim}
\usepackage{amsfonts}
\usepackage{hyperref}
\usepackage{cancel}

\begin{document}

\title{Topological diagrams of $\Omega^0_c$ decays in the $SU(3)_F$ limit}

\author{Ying-Xin Lai$^{1}$}
\author{Di Wang$^{1}$}\email{wangdi@hunnu.edu.cn}

\address{
$^1$Department of Physics, Hunan Normal University, Changsha, 410081, China
}

\begin{abstract}
The $\Omega^0_c$ baryon is unique among charmed sextet baryons because it decays through the weak interaction.
In this work, we investigate the topological amplitudes of $\Omega_c^0$ decays in the $SU(3)_F$ limit.
The complete set of tree- and penguin-induced diagrams contributing to $\Omega_c^0$ decays into decuplet and octet baryons is presented.
The linear relations between the topological amplitudes and the $SU(3)$ irreducible amplitudes are derived through tensor analysis.
Several isospin relations are obtained, and additional relations are derived to test the K\"orner-Pati-Woo theorem.

\end{abstract}

\maketitle

\section{Introduction}

Charmed baryon decays play an important role in studies of nonperturbative baryonic transitions.
The $\Omega^0_c$ baryon, as the only sextet baryon that does not undergo strong decays, provides a complementary counterpart to the antitriplet baryons $(\Xi^0_c,\Xi^+_c,\Lambda^+_c)$.
The $\Omega^0_c$ baryon exhibits favorable symmetry properties.
Some decay channels, such as $\Omega^0_c\to \Omega^-\pi^+$, receive contributions only from a single color-favored emission diagram, making them ideal candidates for investigating baryon decay dynamics at the charm scale.
Compared with the $\mathcal{B}_{c\overline 3}\to \mathcal{B}_{10}M$ decays, the
$\Omega_c^0\to \mathcal{B}_{10}M$ decays allow us to extract the ratios of the annihilation amplitudes to the emission amplitude from experimental data, thereby providing further dynamical information on charmed baryon decays.
In recent years, experimental collaborations such as ALICE \cite{ALICE:2023sgl,ALICE:2022cop}, Belle \cite{Belle:2022yaq,Belle:2021dgc,Belle:2017szm}, and LHCb \cite{LHCb:2023fvd} have measured $\Omega^0_c$ decays, providing an opportunity to extract information on the underlying decay dynamics of the $\Omega^0_c$ baryon.
From a theoretical perspective, QCD-inspired approaches, including QCD factorization (QCDF) \cite{Beneke:1999br,Beneke:2000ry,Beneke:2003zv,Beneke:2001ev}, the perturbative QCD approach (PQCD) \cite{Keum:2000ph,Keum:2000wi,Lu:2000em,Lu:2000hj}, and soft-collinear effective theory (SCET) \cite{Bauer:2001cu,Bauer:2001yt}, do not work well for charmed hadron decays because the expansion parameters $\alpha_s(m_c)$ and $\Lambda_{\rm QCD}/m_c$ are not sufficiently small.
The decays of $\Omega_c^0$ have been extensively studied in dynamical models, including the factorization and pole model \cite{Cheng:1996cs,Xu:1992sw,Cheng:1993gf,Hu:2020nkg,Dhir:2015tja}, various quark models \cite{Perez-Marcial:1989sch,Zhang:2025pde,Zeng:2024yiv,Huang:2021ots,Wang:2022zja,Hsiao:2020gtc,Pervin:2006ie,Zhao:2018zcb,Gutsche:2018utw}, and QCD (light-cone) sum rules \cite{Duan:2025aur,Shi:2025ocl,Aliev:2022gxi}.

Apart from model calculations, it is also useful to investigate $\Omega^0_c$ decays within the framework of flavor $SU(3)$ symmetry.
The topological diagram approach is a useful tool for studying heavy-hadron weak decays based on flavor $SU(3)$ symmetry \cite{Rizzo:1980yh,Zeppenfeld:1980ex,Chau:1982da,Chau:1986du,Chau:1987tk,Chau:1990ay}.
It provides a framework for both model-independent analyses of experimental data and model calculations.
Topological diagrams have already been applied to the study of $\Omega^0_c$ decays \cite{Wang:2023uea}.
However, some important issues, such as the complete decomposition of the topological diagrams for $\Omega^0_c$ decays and the corresponding linear relations, remain unresolved.
Because Fermi statistics require the overall wave function of the three-quark state to be antisymmetric, the flavor and spin wave functions of baryons are often coupled.
Consequently, the topological diagrams for baryon decays fall into two or more distinct sets owing to the different spin wave functions.
In Refs.~\cite{Liu:2025hbf,Wang:2024ztg,Wang:2025bdl}, we developed a theoretical framework to analyze the topological amplitudes of heavy baryon decays in the $SU(3)_F$ limit.
In this work, we extend this framework to $\Omega_c^0$ decays.
We analyze the topological amplitudes for the $\Omega_c^0\to \mathcal{B}_{10}M$ and $\Omega_c^0\to \mathcal{B}_{8}M$ decays in the $SU(3)_F$ limit.
The relations between the topological amplitudes and the $SU(3)$ irreducible amplitudes are derived.
In addition, we perform phenomenological analyses of $\Omega_c^0$ decays based on topological amplitudes.

The rest of this paper is organized as follows.
In Sec. \ref{td}, we present the topological diagrams contributing to the $\Omega^0_c\to \mathcal{B}_{10}M$ and $\Omega^0_c\to \mathcal{B}_{8}M$ decays and derive the linear relations among them.
The phenomenological analysis is presented in Sec. \ref{pa}.
Sec. \ref{summary} provides a brief summary.

\section{Topological amplitudes}\label{td}

\subsection{$\Omega_c^0\to \mathcal{B}_{10}M$ decays}

The effective Hamiltonian for charm-quark decays can be written as \cite{Buchalla:1995vs}
 \begin{equation}\label{hsm}
 \mathcal H_{\rm eff}={G_F\over \sqrt 2}
 \left[\sum_{q=d,s}V_{cq_1}^*V_{uq_2}\left(\sum_{q=1}^2C_i(\mu)\mathcal{O}_i(\mu)\right)
 -V_{cb}^*V_{ub}\left(\sum_{i=3}^6C_i(\mu)\mathcal{O}_i(\mu)+C_{8g}(\mu)\mathcal{O}_{8g}(\mu)\right)\right],
 \end{equation}
 where $G_F$ is the Fermi coupling constant, and $C_{i}(\mu)$ are the Wilson coefficients of the operators $\mathcal{O}_i(\mu)$.
The magnetic penguin contributions can be incorporated into the Wilson coefficients of the penguin operators through the following substitutions
\cite{Beneke:1999br}
\begin{eqnarray}
C_{3,5}(\mu)\to& C_{3,5}(\mu) + \frac{\alpha_s(\mu)}{8\pi N_c}
\frac{2m_c^2}{\langle l^2\rangle}C_{8g}^{\rm eff}(\mu),\qquad
C_{4,6}(\mu)\to& C_{4,6}(\mu) - \frac{\alpha_s(\mu)}{8\pi }
\frac{2m_c^2}{\langle l^2\rangle}C_{8g}^{\rm eff}(\mu),\label{mag}
\end{eqnarray}
with the effective Wilson coefficient $C_{8g}^{\rm eff}=C_{8g}+C_5$.
Charm quark decays can be classified into three categories according to their Cabibbo suppression: Cabibbo-favored (CF), singly Cabibbo-suppressed (SCS), and doubly Cabibbo-suppressed (DCS) decays. These categories have the following flavor structures:
\begin{align}
  c\to s\bar d u, \qquad   c\to d\bar du/s\bar su,\qquad c\to d\bar s u,
\end{align}
respectively.
Within the $SU(3)_F$ framework, the weak Hamiltonian for charm decays can be written as \cite{Wang:2020gmn}
 \begin{equation}\label{h}
 \mathcal H_{\rm eff}= \sum_p \sum_{i,j,k=1}^3 (H^{(p)})_{ij}^{k}\mathcal{O}_{ij}^{(p)k},
 \end{equation}
in which
\begin{equation}\label{a5}
\mathcal{O}_{ij}^{(p)k} = \frac{G_F}{\sqrt{2}} \sum_{\rm color} \sum_{\rm current}C_p(\overline q_iq_k)(\overline q_jc).
\end{equation}
Here, $\mathcal{O}_{ij}^{(p)k}$ denotes the four-quark operator carrying the Fermi constant $G_F$ and the Wilson coefficient $C_p$.
The superscript $p$ on $\mathcal{O}_{ij}^{(p)k}$ denotes the perturbative order in the effective theory.
In the Standard Model, $p=0$ and $p=1$ correspond to tree and penguin operators, respectively.
For each value of $p$, there are $27$ distinct operators $ \mathcal{O}^{(p)k}_{ij}$.
The coefficient matrix $H^{(p)}$ is a $3\times 3\times 3$ matrix.
The color indices and current structures of the four-quark operators are summed over because, once the flavor structure is fixed, operators with different color and Dirac structures always appear together.
The coefficients $(H^{(p)})_{ij}^k$ are obtained from the following mappings: $(\bar uq_1)(\bar q_2c)\rightarrow V^*_{cq_2}V_{uq_1}$ for current-current operators and $(\bar qq)(\bar uc)\rightarrow -V^*_{cb}V_{ub}$ for penguin operators.
The nonzero $(H^{(0)})_{ij}^k$ induced by tree operators are
\begin{align}\label{ckm1}
 &(H^{(0)})_{13}^2 = V_{cs}^*V_{ud},  \qquad (H^{(0)})^{2}_{12}=V_{cd}^*V_{ud},\qquad (H^{(0)})^{3}_{13}= V_{cs}^*V_{us}, \qquad (H^{(0)})^{3}_{12}=V_{cd}^*V_{us}.
\end{align}
 The nonzero components $(H^{(1)})_{ij}^k$ induced by the penguin operators are
\begin{align}\label{ckm2}
 &(H^{(1)})_{11}^1 = -V_{cb}^*V_{ub}, \qquad (H^{(1)})_{21}^2=-V_{cb}^*V_{ub}, \qquad (H^{(1)})_{31}^3=-V_{cb}^*V_{ub}.
\end{align}

\begin{figure}
  \centering
  \includegraphics[width=14cm]{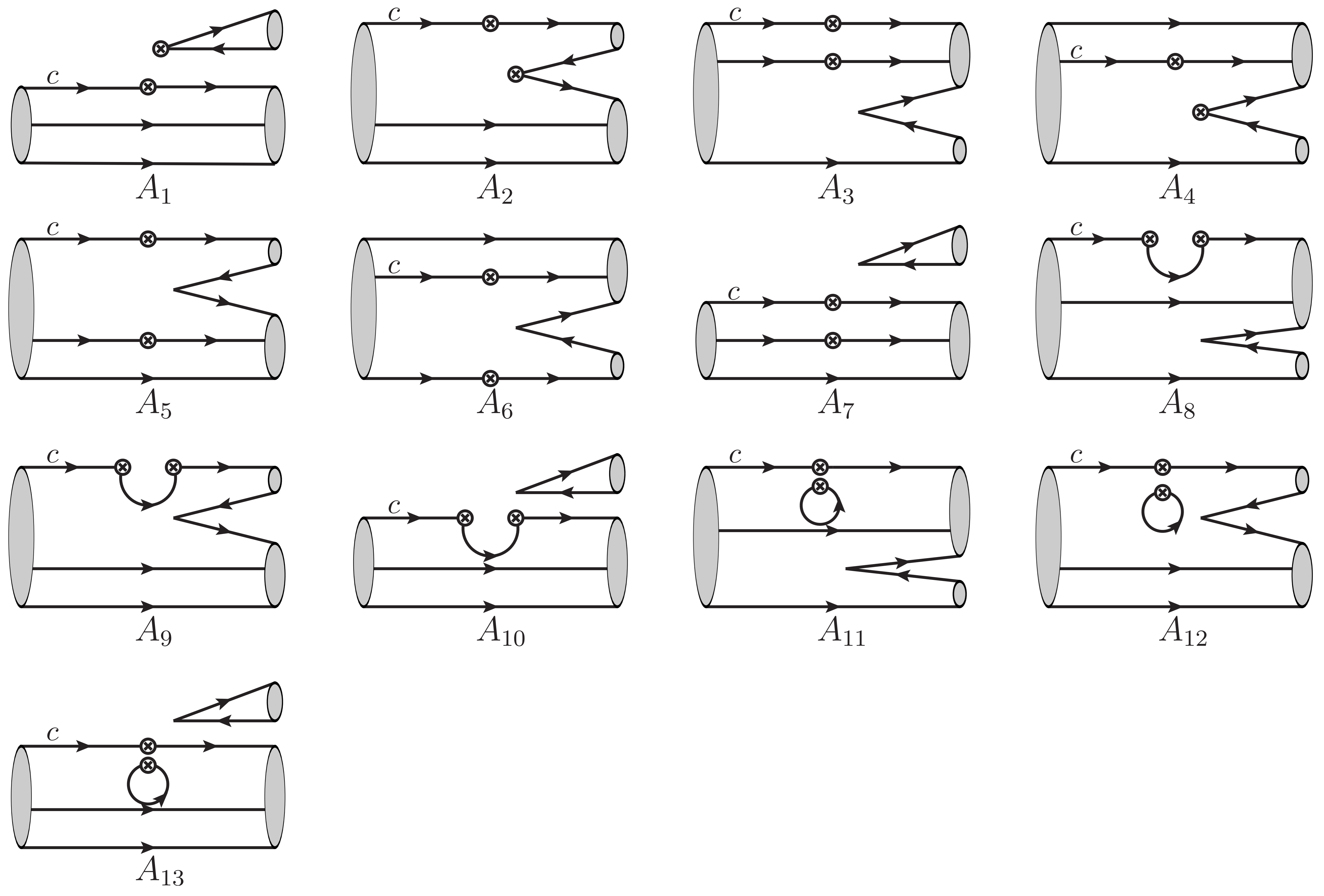}
  \caption{Topological diagrams contributing to the decays of charmed sextet baryons into a decuplet baryon and a pseudoscalar meson.}\label{top6}
\end{figure}

The charmed baryon sextet is expressed as
\begin{eqnarray}
 \mathcal{B}_{c6}=  \left( \begin{array}{ccc}
   \Sigma_c^{++}   &  \frac{1}{\sqrt{2}}\Sigma_c^{+}  & \frac{1}{\sqrt{2}}\Xi_c^{*+} \\
   \frac{1}{\sqrt{2}}\Sigma_c^{+} &   \Sigma_c^{0}   & \frac{1}{\sqrt{2}}\Xi_c^{*0} \\
    \frac{1}{\sqrt{2}}\Xi_c^{*+} & \frac{1}{\sqrt{2}}\Xi_c^{*0} & \Omega_c^0 \\
  \end{array}\right).
\end{eqnarray}
The light pseudoscalar meson nonet is
\begin{eqnarray}
 M=  \left( \begin{array}{ccc}
   \frac{1}{\sqrt 2} \pi^0+  \frac{1}{\sqrt 6} \eta_8    & \pi^+  & K^+ \\
    \pi^- &   - \frac{1}{\sqrt 2} \pi^0+ \frac{1}{\sqrt 6} \eta_8   & K^0 \\
    K^- & \overline K^0 & -\sqrt{2/3}\eta_8 \\
  \end{array}\right) +  \frac{1}{\sqrt 3} \left( \begin{array}{ccc}
   \eta_1    & 0  & 0 \\
    0 &  \eta_1   & 0 \\
   0 & 0 & \eta_1 \\
  \end{array}\right).
\end{eqnarray}
The pseudoscalar mesons $\eta_8$ and $\eta_1$ are not mass eigenstates.
Instead, the mass eigenstates $\eta$ and $\eta^\prime$ are mixtures of $\eta_8$ and $\eta_1$,
\begin{eqnarray}
\left( \begin{array}{ccc}
\eta\\
\eta^\prime
\end{array}
\right)
=
\left(
\begin{array}{cc}
\cos\xi  &  -\sin\xi\\
\sin\xi  &  \cos\xi
\end{array}
\right)\left(
\begin{array}{c}
\eta_8\\
\eta_1
\end{array}\right).
\end{eqnarray}
The light baryons form an $SU(3)$ octet and decuplet.
The light baryon decuplet is given by
\begin{align}
 &\Delta^{++}  = \mathcal{B}_{10}^{111},  \qquad \Delta^{-}=\mathcal{B}_{10}^{222},\qquad  \Omega^-=\mathcal{B}_{10}^{333}, \qquad\Sigma^{*0}=\frac{1}{\sqrt{6}}(\mathcal{B}_{10}^{123} + \mathcal{B}_{10}^{132} +\mathcal{B}_{10}^{213}+ \mathcal{B}_{10}^{231} + \mathcal{B}_{10}^{312} +\mathcal{B}_{10}^{321}),\nonumber\\
 & \Delta^{+} = \frac{1}{\sqrt{3}}(\mathcal{B}_{10}^{112} + \mathcal{B}_{10}^{121} + \mathcal{B}_{10}^{211}),\qquad  \Delta^{0} = \frac{1}{\sqrt{3}}(\mathcal{B}_{10}^{122} + \mathcal{B}_{10}^{212} + \mathcal{B}_{10}^{221}),\qquad \Sigma^{*+}= \frac{1}{\sqrt{3}}(\mathcal{B}_{10}^{113} +\mathcal{B}_{10}^{131} +\mathcal{B}_{10}^{311}), \nonumber\\
&\Sigma^{*-}=\frac{1}{\sqrt{3}} (\mathcal{B}_{10}^{223} +\mathcal{B}_{10}^{232} +\mathcal{B}_{10}^{322}), \qquad
 \Xi^{*0}=\frac{1}{\sqrt{3}}(\mathcal{B}_{10}^{133} +\mathcal{B}_{10}^{313} +\mathcal{B}_{10}^{331}),\qquad
 \Xi^{*-}=\frac{1}{\sqrt{3}}(\mathcal{B}_{10}^{233} + \mathcal{B}_{10}^{323}+\mathcal{B}_{10}^{332}).
\end{align}
The amplitude for charmed sextet baryon decays into a decuplet baryon and a pseudoscalar meson is
\begin{align}\label{am2}
  \mathcal{A}(\mathcal{B}_{c6}\to \mathcal{B}_{10}M) &=A_1(\mathcal{B}_{c6})_{ij}H^k_{lm}M^l_k \mathcal{B}_{10}^{ijm}+A_2(\mathcal{B}_{c6})_{ij}H^k_{lm}M^m_k \mathcal{B}_{10}^{ijl}  +A_3(\mathcal{B}_{c6})_{ij}H^j_{kl}M^i_m\mathcal{B}_{10}^{klm}
 \nonumber\\ &+A_4(\mathcal{B}_{c6})_{ij}H^k_{lm}M^j_k\mathcal{B}_{10}^{ilm}
  +A_5(\mathcal{B}_{c6})_{ij}H^j_{kl}M^l_m \mathcal{B}_{10}^{ikm}
 +  A_6(\mathcal{B}_{c6})_{ij}H^j_{kl}M^k_m \mathcal{B}_{10}^{ilm}\nonumber\\
 & +A_7(\mathcal{B}_{c6})_{ij}H^j_{kl}M^m_m \mathcal{B}_{10}^{ikl}
  +A_8(\mathcal{B}_{c6})_{ij}H^l_{kl}M^j_m\mathcal{B}_{10}^{ikm}
  +A_{9}(\mathcal{B}_{c6})_{ij}H^l_{kl}M^k_m\mathcal{B}_{10}^{ijm}
  \nonumber\\
 &+A_{10}(\mathcal{B}_{c6})_{ij}H^l_{kl}M^m_m\mathcal{B}_{10}^{ijk}
 +A_{11}(\mathcal{B}_{c6})_{ij}H^l_{lk}M^j_m\mathcal{B}_{10}^{ikm}
 +A_{12}(\mathcal{B}_{c6})_{ij}H^l_{lk}M^k_m\mathcal{B}_{10}^{ijm}
 \nonumber\\
 & +A_{13}(\mathcal{B}_{c6})_{ij}H^l_{lk}M^m_m\mathcal{B}_{10}^{ijk}.
\end{align}
Each term in Eq.~\eqref{am2} corresponds to a topological diagram, as shown in Fig.~\ref{top6}.
The first seven diagrams in Fig.~\ref{top6} do not involve quark loops, whereas the last six represent quark-loop contributions.
The tree-induced and penguin-induced amplitudes can be obtained by inserting the tree and penguin operators into these diagrams, respectively.
In this work, the penguin-induced amplitudes are denoted by a superscript $P$ to distinguish them from the tree-induced amplitudes.
The topological amplitudes for the $\Omega^0_c\to \mathcal{B}_{10}M$ decays are presented in Table~\ref{amp8}.
In total, 10 tree-induced diagrams ($A_1\sim A_{10}$) and 13 penguin-induced diagrams ($A_1^P\sim A_{13}^P$) contribute to $\mathcal{B}_{c6}\to \mathcal{B}_{10}M$ decays.

The topological diagrams listed in Eq.~\eqref{am2} form a complete set.
This conclusion follows from the following counting argument.
For $\mathcal{B}_{c6}\to \mathcal{B}_{10}M$ decays, each term in the amplitude contains five covariant and contravariant indices.
Thus, the total number of possible contractions is $5!=120$.
Because the charmed sextet baryon has two symmetric indices ($\mathcal{B}_{c6}^{ij} = \mathcal{B}_{c6}^{ji}$) and the decuplet baryon has three symmetric indices ($\mathcal{B}_{10}^{ijk} = \mathcal{B}_{10}^{jik}=\mathcal{B}_{10}^{ikj} = \mathcal{B}_{10}^{jki}=\mathcal{B}_{10}^{kij} = \mathcal{B}_{10}^{kji}$), the terms $A_3$, $A_4$, $A_5$, $A_6$, $A_7$, $A_8$, and $A_{11}$ in Eq.~\eqref{am2} each correspond to 12 permutations, whereas the terms $A_1$, $A_2$, $A_{9}$, $A_{10}$, $A_{12}$, and $A_{13}$ each correspond to 6 permutations.
The total number of permutations is therefore $7\times 12 + 6\times 6 = 120$.

\begin{table*}
\caption{Topological amplitudes for $\Omega^0_c\to \mathcal{B}_{10}M$ decays.}\label{amp8}
 \small
\begin{tabular}{|c|c|}
\hline\hline
 channel & amplitude \\\hline
 $\Omega^0_c\to \Xi^{* 0}\overline K^0$& $\frac{1}{\sqrt{3}}\lambda_1(A_2+A_4)$ \\\hline
 $\Omega^0_c\to \Omega^{-}\pi^+$& $\lambda_1A_1$ \\\toprule[1.2pt]
 $\Omega^0_c\to \Sigma^{* +}K^-$& $\frac{1}{\sqrt{3}}\lambda_d A_8+\frac{1}{\sqrt{3}}\lambda_s(A_3+A_5+A_8)-\frac{1}{\sqrt{3}}
 \lambda_b(A^P_3+A^P_4+A^P_{6}+A^P_{8}+3A^P_{11})$
 \\\hline
 $\Omega^0_c\to \Sigma^{* 0}\overline K^0$& $\frac{1}{\sqrt{6}}\lambda_d(A_4+A_8)+\frac{1}{\sqrt{6}}\lambda_s(A_3+A_5+A_8)
 -\frac{1}{\sqrt{6}}\lambda_b(A^P_3+A^P_4+A^P_6+A^P_8+3 A^P_{11})$ \\\hline
 $\Omega^0_c\to \Xi^{* 0}\pi^0$& $\frac{1}{\sqrt{6}}\lambda_d(-A_2+A_9)+\frac{1}{\sqrt{6}}\lambda_s(A_6+A_9)
 -\frac{1}{\sqrt{6}}\lambda_b(A_2^P+A_5^P+A_9^P+3 A_{12}^P)$ \\\hline
 $\Omega^0_c\to \Xi^{* 0}\eta_8$&$\frac{1}{3\sqrt{2}}\lambda_d(A_2-2 A_8+A_9)+\frac{1}{3\sqrt{2}}\lambda_s(-2 A_2-2 A_3-2 A_4-2 A_5+A_6-2 A_8+A_9)$ \\ & $-\frac{1}{3\sqrt{2}}\lambda_b(A^P_2-2 A^P_3-2 A^P_4+A^P_5-2 A^P_6-2 A^P_8+A^P_9-6 A^P_{11}+3 A^P_{12})$ \\\hline
 $\Omega^0_c\to \Xi^{* 0}\eta_1$&$\frac{1}{3}\lambda_d(A_2+A_8+A_9+3 A_{10})+\frac{1}{3}\lambda_s(A_2+A_3+A_4+A_5+A_6+3 A_7+A_8+A_9+3 A_{10})$ \\ & $-\frac{1}{3}\lambda_b(3 A^P_1+A^P_2+A^P_3+A^P_4+A^P_5+A^P_6+3 A^P_7+A^P_8+A^P_9+3 A^P_{10}+3 A^P_{11}+3 A^P_{12}+9 A^P_{13})$ \\\hline
 $\Omega^0_c\to \Xi^{* -}\pi^+$&  $\frac{1}{\sqrt{3}}\lambda_d(A_1+A_9)+\frac{1}{\sqrt{3}}\lambda_s(A_6+A_9)
 -\frac{1}{\sqrt{3}}\lambda_b(A^P_2+A^P_5+A^P_9+3 A^P_{12})$ \\\hline
 $\Omega^0_c\to \Omega^{-}K^+$&$\lambda_dA_{9}+\lambda_s(A_1+A_6+A_9)
 -\lambda_b(A^P_2+A^P_5+A^P_9+3 A^P_{12})$  \\\toprule[1.2pt]
 $\Omega^0_c\to \Delta^{ +}K^-$&$\frac{1}{\sqrt{3}}\lambda_2A_3$  \\\hline
 $\Omega^0_c\to \Delta^{ 0}\overline K^0$& $\frac{1}{\sqrt{3}}\lambda_2A_3$ \\\hline
 $\Omega^0_c\to \Sigma^{* +}\pi^-$& $\frac{1}{\sqrt{3}}\lambda_2A_5$ \\\hline
 $\Omega^0_c\to \Sigma^{* 0}\pi^0$& $\frac{1}{2\sqrt{3}}\lambda_2(-A_5+A_6)$ \\\hline
 $\Omega^0_c\to \Sigma^{* 0}\eta_8$& $\frac{1}{6}\lambda_2(-2A_3-2A_4+A_5+A_{6})$ \\\hline
 $\Omega^0_c\to \Sigma^{* 0}\eta_1$&$\frac{1}{3\sqrt{2}}\lambda_2(A_3+A_4+A_5+A_6+3 A_7)$  \\\hline
 $\Omega^0_c\to \Sigma^{* -}\pi^+$&$\frac{1}{\sqrt{3}}\lambda_2A_6$  \\\hline
 $\Omega^0_c\to \Xi^{* 0}K^0$& $\frac{1}{\sqrt{3}}\lambda_2(A_2+A_5)$ \\\hline
 $\Omega^0_c\to \Xi^{* -}K^+$&$\frac{1}{\sqrt{3}}\lambda_2(A_1+A_6)$  \\\hline
  \hline
\end{tabular}
\end{table*}

In the $SU(3)$ framework, the weak operator $\mathcal{O}^{k}_{ij}$ is decomposed as
\begin{align}\label{su}
  \mathcal{O}^k_{ij}= &\mathcal{O}(15)^k_{ij}+\epsilon_{ijl}\mathcal{O}(\overline 6)^{lk}+\delta_j^k\Big(\frac{3}{8}\mathcal{O}( 3)_i-\frac{1}{8}\mathcal{O}(3^\prime)_i\Big)+
  \delta_i^k\Big(\frac{3}{8}\mathcal{O}( 3^\prime)_j-\frac{1}{8}\mathcal{O}( 3)_j\Big).
\end{align}
The nonzero coefficients induced by tree operators in the irreducible representations of $SU(3)$ are given by
\begin{align}\label{ckm3}
 &  H^{(0)}( \overline6)^{22}=-\frac{V_{cs}^*V_{ud}}{2},\qquad H^{(0)}( \overline 6)^{23}=\frac{V_{cd}^*V_{ud}-V_{cs}^*V_{us}}{4},  \qquad H^{(0)}( \overline 6)^{33}=  \frac{V_{cd}^*V_{us}}{2},\nonumber \\
   &  H^{(0)}(15)^{1}_{11}=-\frac{V_{cd}^*V_{ud}+V_{cs}^*V_{us}}{4}, \qquad H^{(0)}(15)^{2}_{13}= \frac{V_{cs}^*V_{ud}}{2},  \qquad  H^{(0)}(15)^{3}_{12}=\frac{V_{cd}^*V_{us}}{2},\nonumber \\
 &  H^{(0)}(15)^{2}_{12}= \frac{3V_{cd}^*V_{ud}-V_{cs}^*V_{us}}{8},\qquad H^{(0)}(15)^{3}_{13}=\frac{3V_{cs}^*V_{us}-V_{cd}^*V_{ud}}{8}, \nonumber \\ & H^{(0)}( 3)_1=V_{cd}^*V_{ud}+V_{cs}^*V_{us}.
\end{align}
The nonzero coefficients induced by penguin operators in the $SU(3)$ irreducible representations are given by
\begin{align}\label{ckm4}
 H^{(1)}( 3)_1=-V_{cb}^*V_{ub}, \qquad H^{(1)}( 3^\prime)_1=-3V_{cb}^*V_{ub}.
\end{align}
The $SU(3)$ irreducible amplitude for the $\mathcal{B}_{c6}\to \mathcal{B}_{10} M$ decay can be expressed as
\begin{align}
  \mathcal{A}^{IR}(\mathcal{B}_{c6}\to \mathcal{B}_{10}M) &= a^{15}_1(\mathcal{B}_{c6})_{ij}H(15)^k_{lm}M^l_k\mathcal{B}_{10}^{ijm}
  +a^{\overline6}_1(\mathcal{B}_{c6})_{ij}H(\overline 6)^k_{lm}M^l_k\mathcal{B}_{10}^{ijm}+ a^{15}_2(\mathcal{B}_{c6})_{ij}H(15)^j_{kl}M^i_m\mathcal{B}_{10}^{klm}\nonumber\\
  &+a^{15}_3(\mathcal{B}_{c6})_{ij}H(15)^k_{lm}M^j_k\mathcal{B}_{10}^{ilm}
  +a^{15}_4(\mathcal{B}_{c6})_{ij}H(15)^j_{kl}M^l_m\mathcal{B}_{10}^{ikm}
+  a^{\overline6}_2(\mathcal{B}_{c6})_{ij}H(\overline 6)^j_{kl}M^l_m\mathcal{B}_{10}^{ikm}\nonumber\\&
+a^{15}_5(\mathcal{B}_{c6})_{ij}H(15)^j_{kl}M^m_m\mathcal{B}_{10}^{ikl}
  +a^{3}_1(\mathcal{B}_{c6})_{ij}H(3)_{k}M^j_m\mathcal{B}_{10}^{ikm}
  +a^{3}_{2}(\mathcal{B}_{c6})_{ij}H(3)_{k}M^k_m\mathcal{B}_{10}^{ijm}\nonumber\\
&+a^{3}_{3}(\mathcal{B}_{c6})_{ij}H(3)_{k}M^m_m\mathcal{B}_{10}^{ijk}
  +a^{3^\prime}_{1}(\mathcal{B}_{c6})_{ij}H(3^\prime)_{k}M^j_m\mathcal{B}_{10}^{ikm}
 +a^{3^\prime}_{2}(\mathcal{B}_{c6})_{ij}H(3^\prime)_{k}M^k_m\mathcal{B}_{10}^{ijm}
 \nonumber\\
 & +a^{3^\prime}_{3}(\mathcal{B}_{c6})_{ij}H(3^\prime)_{k}M^m_m\mathcal{B}_{10}^{ijk}.
\end{align}
The relations between the $SU(3)$ irreducible amplitudes and the topological amplitudes for $\mathcal{B}_{c6}\to \mathcal{B}_{10} M$ decays are derived as
\begin{align}\label{sol}
   & a^{15}_1 = A_1 + A_2,  \quad a^{\overline 6}_1 = A_1 - A_2, \quad a^{15}_2 = A_3,\quad a^{15}_3 = A_4,  \quad a^{15}_4 = A_5 + A_6,  \quad a^{\overline 6}_2 = A_5 - A_6, \quad a^{15}_5 = A_7,  \nonumber\\
   &  a^{3}_1 = \frac{1}{4}A_3+ \frac{1}{4}A_4 + \frac{3}{8}A_5- \frac{1}{8}A_6+A_8,\qquad a^{3^\prime}_{1} = \frac{1}{4}A_3+ \frac{1}{4}A_4 - \frac{1}{8}A_5+ \frac{3}{8}A_6+A_{11}, \nonumber\\
    &  a^{3}_{2} =  \frac{3}{8}A_1- \frac{1}{8}A_2-\frac{1}{8}A_5+ \frac{3}{8}A_6 +A_{9},\qquad a^{3^\prime}_{2} = - \frac{1}{8}A_1+\frac{3}{8}A_2+\frac{3}{8}A_5- \frac{1}{8}A_6 +A_{12},\nonumber\\
    &  a^{3}_{3} =- \frac{1}{8}A_1 + \frac{3}{8}A_2+ \frac{1}{4}A_7+A_{10},\qquad a^{3^\prime}_{3} =\frac{3}{8}A_1 - \frac{1}{8}A_2+\frac{1}{4}A_7+A_{13}.
\end{align}

Note that the $3$-dimensional representation $3^\prime$ is absent in Eq.~\eqref{ckm3}, and the $15$- and $6$-dimensional representations are absent in Eq.~\eqref{ckm4}.
Moreover, the coefficient matrices of the $3$-dimensional representations, including $H^{(0)}(3)$,  $H^{(1)}(3)$, and $H^{(1)}(3^\prime)$, contain only the first component.
Owing to the unitarity of the CKM matrix, we have $H^{(0)}(3)_1 = V_{cd}^*V_{ud}+V_{cs}^*V_{us} = -V_{cb}^*V_{ub}$, and thus
$H^{(0)}(3)_1:H^{(1)}(3)_1:H^{(1)}(3^\prime)_1 = -V_{cb}^*V_{ub}:-V_{cb}^*V_{ub}:-3V_{cb}^*V_{ub}$.
The $SU(3)$ irreducible amplitudes involving $H^{(0)}(3)$,  $H^{(1)}(3)$, and $H^{(1)}(3^\prime)$ appear in the following fixed combinations:
\begin{align}\label{x1}
  a_1^{3,T+P} &= a^3_1 + a^{3P}_1 + 3a^{3^\prime P}_{1},\qquad a_2^{3,T+P} = a^3_2 + a^{3P}_2 + 3a^{3^\prime P}_{2}, \qquad
  a_3^{3,T+P} = a^3_3 + a^{3P}_3 + 3a^{3^\prime P}_{3}.
\end{align}
According to Eq.~\eqref{x1}, the penguin-induced $SU(3)$ irreducible amplitudes always appear together with the tree-induced $SU(3)$ irreducible amplitudes $a^{3}_{1}$, $a^{3}_2$, and $a^3_{3}$.
Thus, the penguin-induced amplitudes have no independent degrees of freedom.
Substituting Eq.~\eqref{sol} into Eq.~\eqref{x1}, we find that the penguin-induced diagrams $A_{1}^P\sim A_{13}^P$ appear together with the tree-induced diagrams $A_{8}$, $A_9$, and $A_{10}$ in the following fixed combinations:
\begin{align}
  A_8^{T+P} &= A_8 + A^P_3 + A^P_4+ A^P_6 +  A^P_8 +3A^P_{11},\qquad A_9^{T+P} = A_9 + A^P_2 + A^P_5 +  A^P_9 +3A^P_{12}, \nonumber\\
  A_{10}^{T+P} &= A_{10} + A^P_1 + A^P_7+ A^P_{10} + 3A^P_{13}.
\end{align}
Ultimately, 10 independent amplitudes contribute to $\mathcal{B}_{c6}\to \mathcal{B}_{10}M$ decays.
In the $SU(3)_F$ limit, all penguin-induced diagrams and tree-induced diagrams involving a quark loop make negligible contributions to the branching fractions, since $|V_{cb}^*V_{ub}| \ll |V_{cd}^*V_{ud}|$ and $|V_{cs}^*V_{us}|$.
However, they are not negligible in CP asymmetries because of the large weak phase of the CKM matrix element $V_{ub}$.

\subsection{$\Omega_c^0\to \mathcal{B}_{8}M$ decays}

\begin{figure}
  \centering
  \includegraphics[width=14cm]{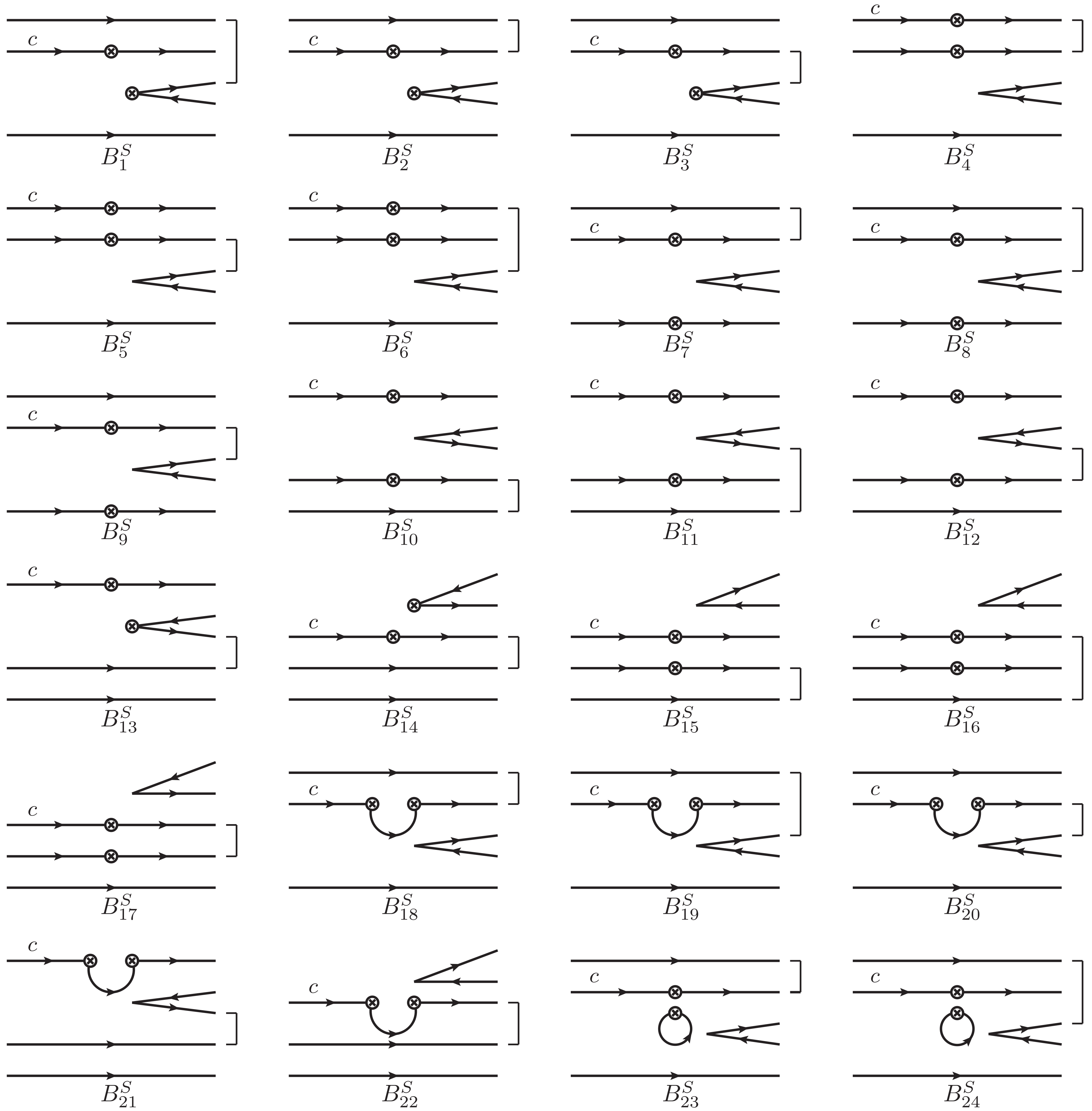}\\
\vspace{1.5mm}
  \includegraphics[width=14cm]{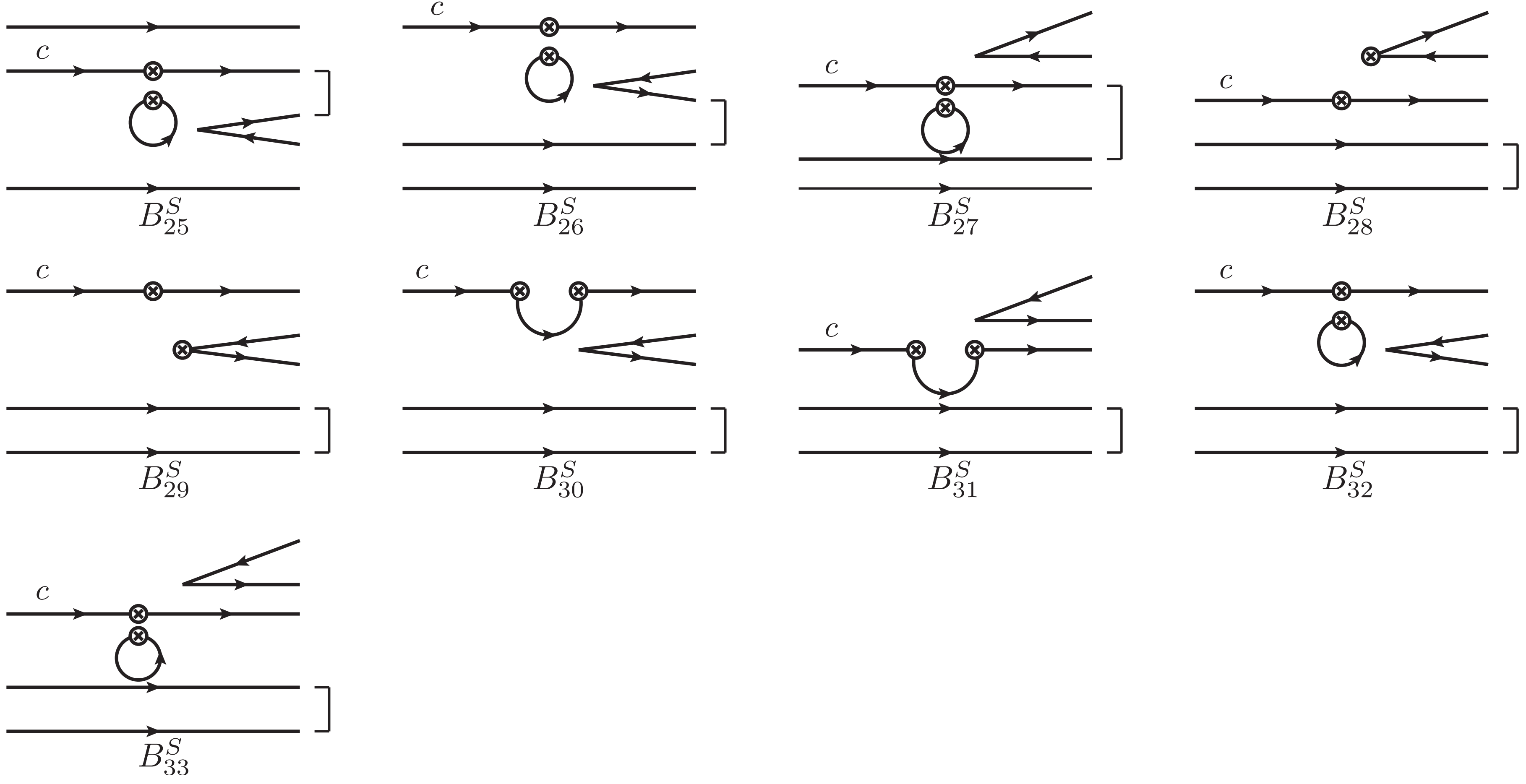}
  \caption{Topological diagrams contributing to $\mathcal{B}_{c6}\to \mathcal{B}_8^SM$ decays, where "$]$" indicates that the two light quarks in the baryon octet are symmetric.}\label{top7}
\end{figure}

\begin{table*}
\caption{Topological amplitudes for $\Omega^0_c\to \mathcal{B}_8^S M$ decays.}\label{amp15}
 \small
\begin{tabular}{|c|c|}
\hline\hline
 channel & amplitude \\\hline
 $\Omega^0_c\to \Xi^{ 0}\overline K^0$& $\frac{1}{\sqrt{6}}\lambda_1(B^S_1-2B^S_2+B^S_3+B^S_{13}-2B^S_{29})$ \\\toprule[1.2pt]
$\Omega^0_c\to \Sigma^{+}K^-$&$\frac{1}{\sqrt{6}}\lambda_d(-B^S_{18}-B^S_{19}+2B^S_{20})
$\\&$+\frac{1}{\sqrt{6}}\lambda_s(-B^S_4+2B^S_5-B^S_6-B^S_{10}-B^S_{11}+2B^S_{12}-B^S_{18}-B^S_{19}+2B^S_{20})$
 \\&
 $-\frac{1}{\sqrt{6}}\lambda_b(-B^{SP}_{1}-B^{SP}_{2}+2B^{SP}_{3}-B^{SP}_{4}-B^{SP}_{5}+2B^{SP}_{6}-B^{SP}_{7}-B^{SP}_{8}
 +2B^{SP}_{9}$\\&$-B^{SP}_{18}-B^{SP}_{19}+2B^{SP}_{20}-3B^{SP}_{23}-3B^{SP}_{24}+6B^{SP}_{25})$ \\\hline
 $\Omega^0_c\to \Sigma^{ 0}\overline K^0$& $\frac{1}{2\sqrt{3}}\lambda_d(B^S_1+B^S_2-2B^S_3+B^S_{18}+B^S_{19}-2B^S_{20})$\\&
$+\frac{1}{2\sqrt{3}}\lambda_s(B^S_4-2B^S_5+B^S_6+B^S_{10}+
 B^S_{11}-2B^S_{12}+B^S_{18}+B^S_{19}-2B^S_{20})$
 \\&
 $-\frac{1}{2\sqrt{3}}\lambda_b(B^{SP}_1+B^{SP}_2-2B^{SP}_3+B^{SP}_4+B^{SP}_5-2B^{SP}_6+B^{SP}_7+B^{SP}_8-2B^{SP}_9$
 \\&$+B^{SP}_{18}+B^{SP}_{19}-2B^{SP}_{20}+3B_{23}^S+3B^{SP}_{24}-6B^{SP}_{25})$ \\\hline
$\Omega^0_c\to \Xi^{ 0}\pi^0$& $\frac{1}{2\sqrt{3}}\lambda_d(-B^S_{13}+2B^S_{29}+B^S_{21}-2B^S_{30})+\frac{1}{2\sqrt{3}}\lambda_s(-2B^S_7+B^S_8+B^S_9+B^S_{21}-2B^S_{30})$
 \\&
 $-\frac{1}{2\sqrt{3}}\lambda_b(-2B^{SP}_{10}+B^{SP}_{11}+B^{SP}_{12}+B^{SP}_{13}-2B^{SP}_{29}+B^{SP}_{21}-2B^{SP}_{30}+3B^{SP}_{26}-6B^{SP}_{32})$ \\\hline
$\Omega^0_c\to \Xi^{ 0}\eta_8$& $\frac{1}{6}\lambda_d(B^S_{13}-2B^S_{29}-2B^S_{18}+4B^S_{19}-2B^S_{20}+B^S_{21}-2B^S_{30})$\\&
$+\frac{1}{6}\lambda_s(-2B^S_1+4B^S_2-2B^S_3-2B^S_{4}-2B^S_{5}+4B^S_6-2B^S_7+B^S_8+B^S_{9}-2B^S_{10}$\\
&$+4B^S_{11}-2B^S_{12}-2B^S_{13}+4B^S_{29}-2B^S_{18}+4B^S_{19}-2B^S_{20}+B^S_{21}-2B^S_{30})$
 \\&
 $-\frac{1}{6}\lambda_b(4B^{SP}_{1}-2B^{SP}_{2}-2B^{SP}_{3}-2B^{SP}_{4}+4B^{SP}_{5}-2B^{SP}_{6}-2B^{SP}_{7}
 +4B^{SP}_{8}-2B^{SP}_{9}$\\&
 $-2B^{SP}_{10}+B^{SP}_{11}+B^{SP}_{12}-2B^{SP}_{18}+4B^{SP}_{19}-2B^{SP}_{20}+B^{SP}_{21}-2B^{SP}_{30}$\\
&$-6B^{SP}_{23}+12B^{SP}_{24}-6B^{SP}_{25}+3B^{SP}_{26}-6B^{SP}_{32})$ \\\hline
$\Omega^0_c\to \Xi^{ 0}\eta_1$&  $\frac{1}{3\sqrt{2}}\lambda_d(B^S_{13}-2B^S_{29}+B^S_{18}-2B^S_{19}+B^S_{20}+B^S_{21}-2B^S_{30}+3B^S_{22}-6B^S_{31})$\\&
$+\frac{1}{3\sqrt{2}}\lambda_s(B^S_1-2B^S_2+B^S_3+B^S_{4}+B^S_{5}-2B^S_6-2B^S_7+B^S_8+B^S_{9}+B^S_{10}-2B^S_{11}
+B^S_{12}$\\
&$+B^S_{13}-2B^S_{29}+3B^S_{15}-6B^S_{16}+3B^S_{17}+B^S_{18}-2B^S_{19}+B^S_{20}+B^S_{21}-2B^S_{30}+3B^S_{22}-6B^S_{31})$
 \\&
 $-\frac{1}{3\sqrt{2}}\lambda_b(-2B^{SP}_{1}+B^{SP}_{2}+B^{SP}_{3}+B^{SP}_{4}-2B^{SP}_{5}+B^{SP}_{6}+B^{SP}_{7}-2B^{SP}_{8}
 +B^{SP}_{9}-2B^{SP}_{10}$\\&
 $+B^{SP}_{11}+B^{SP}_{12}+3B^{SP}_{14}-6B^{SP}_{28}-6B^{SP}_{15}+3B^{SP}_{16}+3B^{SP}_{17}+B^{SP}_{18}
 -2B^{SP}_{19}+B^{SP}_{20}+B^{SP}_{21}$\\&$-2B^{SP}_{30}+3B^{SP}_{22}-6B^{SP}_{31}+3B^{SP}_{23}-6B^{SP}_{24}+3B^{SP}_{25}
 +3B^{SP}_{26}-6B^{SP}_{32}+9B^{SP}_{27}-
 18B^{SP}_{33})$ \\\hline
$\Omega^0_c\to \Xi^{ -}\pi^+$&  $\frac{1}{\sqrt{6}}\lambda_d(-B^S_{14}+2B^S_{28}-B^S_{21}+2B^S_{30})+\frac{1}{\sqrt{6}}\lambda_s(2B^S_7-B^S_8-B^S_9-B^S_{21}+2B^S_{30})$
 \\&
 $-\frac{1}{\sqrt{6}}\lambda_b(2B^{SP}_{10}-B^{SP}_{11}-B^{SP}_{12}-B^{SP}_{13}+2B^{SP}_{29}-B^{SP}_{21}+2B^{SP}_{30}-3B^{SP}_{26}+6B^{SP}_{32})$ \\\hline
$\Omega^0_c\to \Lambda^{ 0}\overline K^0$& $\frac{1}{2}\lambda_d(-B^S_{1}+B^S_{2}-B^S_{18}+B^S_{19})+\frac{1}{2}\lambda_s(-B^S_4+B^S_6-B^S_{10}+B^S_{11}-B^S_{18}+B^S_{19})$
 \\&
 $-\frac{1}{2}\lambda_b(B^{SP}_{1}-B^{SP}_{2}-B^{SP}_{4}+B^{SP}_{5}-B^{SP}_{7}+B^{SP}_{8}-B^{SP}_{18}+B^{SP}_{19}-3B^{SP}_{23}+3B^{SP}_{24})$ \\\toprule[1.2pt]
$\Omega^0_c\to \Sigma^{+}\pi^-$& $\frac{1}{\sqrt{6}}\lambda_2(-B^S_{10}-B^S_{11}+2B^S_{12})$ \\\hline
$\Omega^0_c\to \Sigma^{0}\pi^0$& $\frac{1}{2\sqrt{6}}\lambda_2(B^S_{7}+B^S_{8}-2B^S_9-B^S_{10}-B^S_{11}+2B^S_{12})$ \\\hline
$\Omega^0_c\to \Sigma^{ 0}\eta_8$& $\frac{1}{6\sqrt{2}}\lambda_2(-2B^S_1-2B^S_2+4B^S_3+4B^S_{4}-2B^S_{5}-2B^S_6+B^S_{7}+B^S_{8}-2B^S_9+B^S_{10}+B^S_{11}-2B^S_{12})$ \\\hline
$\Omega^0_c\to \Sigma^{ 0}\eta_1$& $\frac{1}{6}\lambda_2(B^S_1+B^S_2-2B^S_3-2B^S_{4}+B^S_{5}+B^S_6+B^S_{7}+B^S_{8}$\\
&$-2B^S_9+B^S_{10}+B^S_{11}-2B^S_{12}+3B^S_{15}+3B^S_{16}-6B^S_{17})$\\\hline
$\Omega^0_c\to \Sigma^{ -}\pi^+$& $\frac{1}{\sqrt{6}}\lambda_2(B^S_{7}+B^S_{8}-2B^S_{9})$\\\hline
$\Omega^0_c\to pK^-$& $\frac{1}{\sqrt{6}}\lambda_2(B^S_{4}-2B^S_{5}+B^S_{6})$ \\\hline
$\Omega^0_c\to n\overline K^0$& $\frac{1}{\sqrt{6}}\lambda_2(-B^S_{4}-B^S_{5}+2B^S_{6})$ \\\hline
$\Omega^0_c\to \Xi^{ 0}K^0$& $\frac{1}{\sqrt{6}}\lambda_2(B^S_{10}-2B^S_{11}+B^S_{12}+B^S_{13}-2B^S_{29})$ \\\hline
$\Omega^0_c\to \Xi^{ -}K^+$& $\frac{1}{\sqrt{6}}\lambda_2(-B^S_{7}+2B^S_{8}-B^S_{9}-B^S_{14}+2B^S_{28})$ \\\hline
$\Omega^0_c\to \Lambda^{ 0}\pi^0$& $\frac{1}{2\sqrt{2}}\lambda_2(B^S_{7}-B^S_{8}+B^S_{10}-B^S_{11})$ \\\hline
$\Omega^0_c\to \Lambda^{ 0}\eta_8$& $\frac{1}{2\sqrt{6}}\lambda_2(2B^S_{1}-2B^S_{2}+2B^S_{5}-2B^S_{6}+B^S_{7}-B^S_{8}-B^S_{10}+B^S_{11})$ \\\hline
$\Omega^0_c\to \Lambda^{ 0}\eta_1$& $\frac{1}{2\sqrt{3}}\lambda_2(-B^S_{1}+B^S_{2}-B^S_{5}+B^S_{6}+B^S_{7}-B^S_{8}-B^S_{10}+B^S_{11}-3B^S_{15}+3B^S_{16})$ \\\hline
  \hline
\end{tabular}
\end{table*}
\begin{table*}
\caption{Topological amplitudes for $\Omega^0_c\to \mathcal{B}_8^A M$ decays.}\label{amp16}
 \small
\begin{tabular}{|c|c|}
\hline\hline
 channel & amplitude \\\hline
 $\Omega^0_c\to \Xi^{ 0}\overline K^0$& $\frac{1}{\sqrt{2}}\lambda_1(B^A_{1}-B^A_{3}+B^A_{13})$ \\\toprule[1.2pt]
$\Omega^0_c\to \Sigma^{+}K^-$&
$\frac{1}{\sqrt{2}}\lambda_d(B^A_{18}+B^A_{19})+\frac{1}{\sqrt{2}}\lambda_s(-B^A_4+B^A_6+B^A_{10}+B^A_{11}+B^A_{18}+B^A_{19})$\\&
$-\frac{1}{\sqrt{2}}\lambda_b(B^{AP}_{1}+B^{AP}_{2}+B^{AP}_{4}+B^{AP}_{5}+B^{AP}_{7}+B^{AP}_{8}+B^{AP}_{18}+B^{AP}_{19}+3B^{AP}_{23}+3B^{AP}_{24})$ \\\hline
 $\Omega^0_c\to \Sigma^{ 0}\overline K^0$&   $\frac{1}{2}\lambda_d(-B^A_1-B^A_2-B^A_{18}-B^A_{19})+\frac{1}{2}\lambda_s(B^A_4-B^A_6-B^A_{10}-B^A_{11}-B^A_{18}-B^A_{19})$
 \\&
 $-\frac{1}{2}\lambda_b(-B^{AP}_1-B^{AP}_2-B^{AP}_4-B^{AP}_5-B^{AP}_7-B^{AP}_8-B^{AP}_{18}-B^{AP}_{19}-3B^{AP}_{23}-3B^{AP}_{24})$ \\\hline
$\Omega^0_c\to \Xi^{ 0}\pi^0$&   $\frac{1}{2}\lambda_d(-B^A_{13}+B^A_{21})+\frac{1}{2}\lambda_s(B^A_8+B^A_9+B^A_{21})-\frac{1}{2}\lambda_b(B^{AP}_{11}+B^{AP}_{12}+B^{AP}_{13}+B^{AP}_{21}+3B^{AP}_{26})$ \\\hline
$\Omega^0_c\to \Xi^{ 0}\eta_8$&
$\frac{1}{2\sqrt{3}}\lambda_d(B^A_{13}-2B^A_{18}+2B^A_{20}+B^A_{21})+\frac{1}{2\sqrt{3}}\lambda_s(-2B^A_1+2B^A_3+2B^A_{4}$\\ &
$+2B^A_5+B^A_8+B^A_{9}-2B^A_{10}+2B^A_{12}-2B^A_{13}-2B^A_{18}+2B^A_{20}+B^A_{21})$\\&
$-\frac{1}{2\sqrt{3}}\lambda_b(-2B^{AP}_{2}-2B^{AP}_{3}-2B^{AP}_{4}+2B^{AP}_{6}-2B^{AP}_{7}+2B^{AP}_{9}+B^{AP}_{11}+B^{AP}_{12}
$\\&$+B^{AP}_{13}-2B^{AP}_{18}+2B^{AP}_{20}+B^{AP}_{21}-6B^{AP}_{23}+6B^{AP}_{25}+3B^A_{26})$ \\\hline
$\Omega^0_c\to \Xi^{ 0}\eta_1$&
$\frac{1}{2\sqrt{3}}\lambda_d(B^A_{13}+B^A_{18}-B^A_{20}+B^A_{21}+3B^A_{22})+\frac{1}{2\sqrt{3}}\lambda_s(B^A_1-B^A_3-B^A_{4}-B^A_5$\\ &
$+B^A_8+B^A_{9}+B^A_{10}-B^A_{12}+B^A_{13}+3B^A_{15}-3B^A_{17}+B^A_{18}-B^A_{20}+B^A_{21}+3B^A_{22})$\\&
$-\frac{1}{2\sqrt{3}}\lambda_b(B^{AP}_{2}+B^{AP}_{3}+B^{AP}_{4}-B^{AP}_{6}+B^{AP}_{7}-B^{AP}_{9}+B^{AP}_{11}+B^{AP}_{12}+3B^{AP}_{14}$\\&
$+3B^{AP}_{16}+3B^{AP}_{17}+B^{AP}_{18}-B^A_{20}+B^{AP}_{21}+3B^{AP}_{22}+3B^{AP}_{23}-3B^{AP}_{25}+3B^{AP}_{26}+9B^{AP}_{27})$ \\\hline
$\Omega^0_c\to \Xi^{ -}\pi^+$&
$\frac{1}{\sqrt{2}}\lambda_d(-B^A_{14}-B^A_{21})+\frac{1}{\sqrt{2}}\lambda_s(-B^A_8-B^A_9-B^A_{21})$\\&$-\frac{1}{\sqrt{2}}\lambda_b(-B^{AP}_{11}-B^{AP}_{12}-B^{AP}_{13}-B^{AP}_{21}-3B^{AP}_{26})$ \\\hline
$\Omega^0_c\to \Lambda^{ 0}\overline K^0$&  $\frac{1}{2\sqrt{3}}\lambda_d(-B^A_{1}+B^A_{2}+2B^A_{3}-B^A_{18}+B^A_{19}+2B^A_{20})$\\&$+
\frac{1}{2\sqrt{3}}\lambda_s(B^A_4+2B^A_5+B^A_{6}-B^A_{10}+B^A_{11}+2B^A_{12}-B^A_{18}+B^A_{19}+2B^A_{20})$\\&
$-\frac{1}{2\sqrt{3}}\lambda_b(B^{AP}_{1}-B^{AP}_{2}-2B^{AP}_{3}-B^{AP}_{4}+B^{AP}_{5}+2B^{AP}_{6}-B^{AP}_{7}+B^{AP}_{8}+2B^{AP}_{9}
$\\&$-B^{AP}_{18}+B^{AP}_{19}+2B^{AP}_{20}-3B^{AP}_{23}+3B^{AP}_{24}+6B^{AP}_{25})$ \\\toprule[1.2pt]
$\Omega^0_c\to \Sigma^{+}\pi^-$& $\frac{1}{\sqrt{2}}\lambda_2(B^A_{10}+B^A_{11})$ \\\hline
$\Omega^0_c\to \Sigma^{0}\pi^0$& $-\frac{1}{2\sqrt{2}}\lambda_2(B^A_{7}+B^A_{8}-B^A_{10}-B^A_{11})$ \\\hline
$\Omega^0_c\to \Sigma^{ 0}\eta_8$&$-\frac{1}{2\sqrt{6}}\lambda_2(-2B^A_{1}-2B^A_{2}+2B^A_{5}+2B^A_{6}+B^A_{7}+B^A_{8}+B^A_{10}+B^A_{11})$  \\\hline
$\Omega^0_c\to \Sigma^{ -}\pi^+$& $-\frac{1}{\sqrt{2}}\lambda_2(B^A_{7}+B^A_{8})$ \\\hline
$\Omega^0_c\to \Sigma^{ 0}\eta_1$& $-\frac{1}{2\sqrt{3}}\lambda_2(B^A_{1}+B^A_{2}-B^A_{5}-B^A_{6}+B^A_{7}+B^A_{8}+B^A_{10}+B^A_{11}+3B^A_{15}+3B^A_{16})$ \\\hline
$\Omega^0_c\to pK^-$& $\frac{1}{\sqrt{2}}\lambda_2(B^A_{4}-B^A_{6})$ \\\hline
$\Omega^0_c\to n\overline K^0$& $\frac{1}{\sqrt{2}}\lambda_2(B^A_{4}+B^A_{5})$ \\\hline
$\Omega^0_c\to \Xi^{ 0}K^0$& $\frac{1}{\sqrt{2}}\lambda_2(B^A_{10}-B^A_{12}+B^A_{13})$ \\\hline
$\Omega^0_c\to \Xi^{ -}K^+$& $\frac{1}{\sqrt{2}}\lambda_2(-B^A_{7}+B^A_{9}-B^A_{14})$ \\\hline
$\Omega^0_c\to \Lambda^{ 0}\pi^0$& $\frac{1}{2\sqrt{6}}\lambda_2(B^A_{7}-B^A_{8}-2B^A_{9}+B^A_{10}-B^A_{11}-2B^A_{12})$ \\\hline
$\Omega^0_c\to \Lambda^{ 0}\eta_8$& $\frac{1}{6\sqrt{2}}\lambda_2(2B^A_{1}-2B^A_{2}-4B^A_{3}-4B^A_{4}-2B^A_{5}+2B^A_{6}$\\ &
$+B^A_{7}-B^A_{8}-2B^A_{9}-B^A_{10}+B^A_{11}+2B^A_{12})$ \\\hline
$\Omega^0_c\to \Lambda^{ 0}\eta_1$& $\frac{1}{6}\lambda_2(-B^A_{1}+B^A_{2}+2B^A_{3}+2B^A_{4}+B^A_{5}-B^A_{6}+B^A_{7}-B^A_{8}$\\ &
$-2B^A_{9}-B^A_{10}+B^A_{11}+2B^A_{12}-3B^A_{15}+3B^A_{16}+6B^A_{17})$\\\hline
  \hline
\end{tabular}
\end{table*}

There are two octets that are symmetric and antisymmetric under $q_1\leftrightarrow q_2$, denoted by $\mathcal{B}_8^S$ and $\mathcal{B}_8^A$, respectively.
The light baryon octets $\mathcal{B}_8^S$ and $\mathcal{B}_8^A$ can be expressed as tensors with three covariant indices.
$\mathcal{B}_8^S$ is given by
\begin{align}\label{BS}
  \Sigma^+ & = \frac{1}{\sqrt{6}}(2\mathcal{B}_8^{113}-\mathcal{B}_8^{311}-\mathcal{B}_8^{131}),
  \qquad   p =\frac{1}{\sqrt{6}}(-2\mathcal{B}_8^{112}+\mathcal{B}_8^{211}+\mathcal{B}_8^{121}), \qquad   \Sigma^- = \frac{1}{\sqrt{6}}(-2\mathcal{B}_8^{223}+\mathcal{B}_8^{322}+\mathcal{B}_8^{232}), \nonumber\\
    n &=\frac{1}{\sqrt{6}}(2\mathcal{B}_8^{221}-\mathcal{B}_8^{122}-\mathcal{B}_8^{212}), \qquad\Xi^- = \frac{1}{\sqrt{6}}(2\mathcal{B}_8^{332}-\mathcal{B}_8^{233}-\mathcal{B}_8^{323}), \qquad   \Xi^0 =\frac{1}{\sqrt{6}}(-2\mathcal{B}_8^{331}+\mathcal{B}_8^{133}+\mathcal{B}_8^{313}),\nonumber\\
 \Sigma^0 &=\frac{1}{\sqrt{12}}(\mathcal{B}_8^{321}+\mathcal{B}_8^{231}+\mathcal{B}_8^{312}
 +\mathcal{B}_8^{132}-2\mathcal{B}_8^{123}-2\mathcal{B}_8^{213}), \qquad \Lambda^0 = \frac{1}{2}(\mathcal{B}_8^{132}-\mathcal{B}_8^{231}+\mathcal{B}_8^{312}-\mathcal{B}_8^{321}
 ),
\end{align}
and $\mathcal{B}_8^A$ is given by
\begin{align}\label{BA}
  \Sigma^+ & = \frac{1}{\sqrt{2}}(\mathcal{B}_8^{311}-\mathcal{B}_8^{131}),\qquad   p =\frac{1}{\sqrt{2}}(\mathcal{B}_8^{121}-\mathcal{B}_8^{211}), \qquad \Sigma^- = \frac{1}{\sqrt{2}}(\mathcal{B}_8^{232}-\mathcal{B}_8^{322}),  \nonumber\\
  n& =\frac{1}{\sqrt{2}}(\mathcal{B}_8^{122}-\mathcal{B}_8^{212}), \qquad  \Xi^- = \frac{1}{\sqrt{2}}(\mathcal{B}_8^{233}-\mathcal{B}_8^{323}),\qquad   \Xi^0 =\frac{1}{\sqrt{2}}(\mathcal{B}_8^{313}-\mathcal{B}_8^{133}),  \nonumber\\
\Sigma^0 &=\frac{1}{2}(\mathcal{B}_8^{132}-\mathcal{B}_8^{312}+\mathcal{B}_8^{231}
-\mathcal{B}_8^{321}),   \qquad   \Lambda^0  = \frac{1}{\sqrt{12}}(2\mathcal{B}_8^{213}-2\mathcal{B}_8^{123}+\mathcal{B}_8^{231}
-\mathcal{B}_8^{321}+\mathcal{B}_8^{312}-\mathcal{B}_8^{132}).
\end{align}
According to the requirement of quantum statistics that the overall wave function of the three-quark state be antisymmetric, the combined flavor-spin wave function must be symmetric under the interchange of any two quarks.
The mixed-symmetry combinations $\phi_S\chi_S$ and $\phi_A\chi_A$ are both symmetric under the interchange $q_1\leftrightarrow q_2$.
However, neither combination alone has definite symmetry under the interchanges $q_1\leftrightarrow q_3$ and $q_2\leftrightarrow q_3$.
Only the equal-weight linear combination $\phi_S\chi_S+\phi_A\chi_A$ is totally symmetric under the interchange of any two quarks.
Because of the different spin structures of the two octets $\phi_S$ and $\phi_A$, the topological diagrams for charmed baryon decays into octet baryons fall into two distinct sets.
For clarity, we label the topological diagrams for charmed baryon decays into $\mathcal{B}^S_8$ and $\mathcal{B}^A_8$ with the superscripts $S$ and $A$, respectively.
The topological amplitude for $\mathcal{B}_{c6}\to \mathcal{B}_8^SM$ decays is constructed as
\begin{align}\label{amp1}
 \mathcal{A}^S(\mathcal{B}_{c6}\to \mathcal{B}_8^SM) &= B^S_1(\mathcal{B}_{c6})_{ij} H^m_{kl}M^i_m ( \mathcal{B}_8^S)^{jkl} + B^S_2(\mathcal{B}_{c6})_{ij}H^m_{kl}M^i_m ( \mathcal{B}_8^S)^{jlk}+ B^S_3(\mathcal{B}_{c6})_{ij}H^m_{kl}M^i_m ( \mathcal{B}_8^S)^{klj} \nonumber\\
   & + B^S_4(\mathcal{B}_{c6})_{ij} H^i_{kl}M^j_m ( \mathcal{B}_8^S)^{klm}  + B^S_5(\mathcal{B}_{c6})_{ij} H^i_{kl}M^j_m ( \mathcal{B}_8^S)^{kml}  + B^S_6(\mathcal{B}_{c6})_{ij} H^i_{kl}M^j_m ( \mathcal{B}_8^S)^{lmk} \nonumber\\
   &+ B^S_7(\mathcal{B}_{c6})_{ij} H^i_{kl}M^k_m ( \mathcal{B}_8^S)^{jlm}+ B^S_8(\mathcal{B}_{c6})_{ij} H^i_{kl}M^k_m ( \mathcal{B}_8^S)^{jml} + B^S_9(\mathcal{B}_{c6})_{ij} H^i_{kl}M^k_m ( \mathcal{B}_8^S)^{lmj} \nonumber\\
 &+ B^S_{10}(\mathcal{B}_{c6})_{ij} H^i_{kl}M^l_m ( \mathcal{B}_8^S)^{jkm}  + B^S_{11}(\mathcal{B}_{c6})_{ij} H^i_{kl}M^l_m ( \mathcal{B}_8^S)^{jmk} + B^S_{12}(\mathcal{B}_{c6})_{ij} H^i_{kl}M^l_m ( \mathcal{B}_8^S)^{kmj} \nonumber\\
&+ B^S_{13}(\mathcal{B}_{c6})_{ij} H^m_{kl}M^l_m (\mathcal{B}_8^S)^{ikj}+ B^S_{14}(\mathcal{B}_{c6})_{ij} H^m_{kl}M^k_m ( \mathcal{B}_8^S)^{ilj} + B^S_{15}(\mathcal{B}_{c6})_{ij} H^i_{kl}M^m_m ( \mathcal{B}_8^S)^{jkl} \nonumber\\
 &+ B^S_{16}(\mathcal{B}_{c6})_{ij} H^i_{kl}M^m_m (\mathcal{B}_8^S)^{jlk} + B^S_{17}(\mathcal{B}_{c6})_{ij} H^i_{kl}M^m_m ( \mathcal{B}_8^S)^{klj}  + B^S_{18}(\mathcal{B}_{c6})_{ij} H^l_{kl}M^i_m ( \mathcal{B}_8^S)^{jkm}\nonumber\\
 & + B^S_{19}(\mathcal{B}_{c6})_{ij} H^l_{kl}M^i_m (\mathcal{B}_8^S)^{jmk} + B^S_{20}(\mathcal{B}_{c6})_{ij} H^l_{kl}M^i_m (\mathcal{B}_8^S)^{kmj} + B^S_{21}(\mathcal{B}_{c6})_{ij} H^l_{kl}M^k_m ( \mathcal{B}_8^S)^{imj}\nonumber\\
& + B^S_{22}(\mathcal{B}_{c6})_{ij} H^l_{kl}M^m_m (\mathcal{B}_8^S)^{ikj} + B^S_{23}(\mathcal{B}_{c6})_{ij} H^l_{lk}M^i_m ( \mathcal{B}_8^S)^{jkm}+ B^S_{24}(\mathcal{B}_{c6})_{ij} H^l_{lk}M^i_m ( \mathcal{B}_8^S)^{jmk}\nonumber\\
& + B^S_{25}(\mathcal{B}_{c6})_{ij} H^l_{lk}M^i_m (\mathcal{B}_8^S)^{kmj} + B^S_{26}(\mathcal{B}_{c6})_{ij} H^l_{lk}M^k_m ( \mathcal{B}_8^S)^{imj}+ B^S_{27}(\mathcal{B}_{c6})_{ij} H^l_{lk}M^m_m ( \mathcal{B}_8^S)^{ikj}\nonumber\\
&+ B^S_{28}(\mathcal{B}_{c6})_{ij} H^m_{kl}M^k_m (\mathcal{B}_8^S)^{ijl} + B^S_{29}(\mathcal{B}_{c6})_{ij} H^m_{kl}M^l_m (\mathcal{B}_8^S)^{ijk} + B^S_{30}(\mathcal{B}_{c6})_{ij} H^l_{kl}M^k_m ( \mathcal{B}_8^S)^{ijm} \nonumber\\
&+ B^S_{31}(\mathcal{B}_{c6})_{ij} H^l_{kl}M^m_m ( \mathcal{B}_8^S)^{ijk}+ B^S_{32}(\mathcal{B}_{c6})_{ij} H^l_{lk}M^k_m ( \mathcal{B}_8^S)^{ijm}+ B^S_{33}(\mathcal{B}_{c6})_{ij} H^l_{lk}M^m_m ( \mathcal{B}_8^S)^{ijk},
\end{align}
and the topological amplitude for the $\mathcal{B}_{c6}\to \mathcal{B}_8^AM$ decays is constructed as
\begin{align}\label{amp2}
 \mathcal{A}^A(\mathcal{B}_{c6}\to \mathcal{B}_8^AM) &=  B^A_1(\mathcal{B}_{c6})_{ij} H^m_{kl}M^i_m ( \mathcal{B}_8^A)^{jkl} + B^A_2(\mathcal{B}_{c6})_{ij}H^m_{kl}M^i_m ( \mathcal{B}_8^A)^{jlk}+ B^A_3(\mathcal{B}_{c6})_{ij}H^m_{kl}M^i_m ( \mathcal{B}_8^A)^{klj} \nonumber\\
   & + B^A_4(\mathcal{B}_{c6})_{ij} H^i_{kl}M^j_m (\mathcal{B}_8^A)^{klm}  + B^A_5(\mathcal{B}_{c6})_{ij} H^i_{kl}M^j_m (\mathcal{B}_8^A)^{kml}  + B^A_6(\mathcal{B}_{c6})_{ij} H^i_{kl}M^j_m ( \mathcal{B}_8^A)^{lmk} \nonumber\\
   &+ B^A_7(\mathcal{B}_{c6})_{ij} H^i_{kl}M^k_m (\mathcal{B}_8^A)^{jlm}+ B^A_8(\mathcal{B}_{c6})_{ij} H^i_{kl}M^k_m (\mathcal{B}_8^A)^{jml} + B^A_9(\mathcal{B}_{c6})_{ij} H^i_{kl}M^k_m ( \mathcal{B}_8^A)^{lmj} \nonumber\\
 &+ B^A_{10}(\mathcal{B}_{c6})_{ij} H^i_{kl}M^l_m ( \mathcal{B}_8^A)^{jkm}  + B^A_{11}(\mathcal{B}_{c6})_{ij} H^i_{kl}M^l_m (\mathcal{B}_8^A)^{jmk} + B^A_{12}(\mathcal{B}_{c6})_{ij} H^i_{kl}M^l_m ( \mathcal{B}_8^A)^{kmj} \nonumber\\
&+ B^A_{13}(\mathcal{B}_{c6})_{ij} H^m_{kl}M^l_m ( \mathcal{B}_8^A)^{ikj}+ B^A_{14}(\mathcal{B}_{c6})_{ij} H^m_{kl}M^k_m (\mathcal{B}_8^A)^{ilj} + B^A_{15}(\mathcal{B}_{c6})_{ij} H^i_{kl}M^m_m ( \mathcal{B}_8^A)^{jkl} \nonumber\\
 &+ B^A_{16}(\mathcal{B}_{c6})_{ij} H^i_{kl}M^m_m ( \mathcal{B}_8^A)^{jlk} + B^A_{17}(\mathcal{B}_{c6})_{ij} H^i_{kl}M^m_m (\mathcal{B}_8^A)^{klj}  + B^A_{18}(\mathcal{B}_{c6})_{ij} H^l_{kl}M^i_m ( \mathcal{B}_8^A)^{jkm}\nonumber\\
 & + B^A_{19}(\mathcal{B}_{c6})_{ij} H^l_{kl}M^i_m ( \mathcal{B}_8^A)^{jmk} + B^A_{20}(\mathcal{B}_{c6})_{ij} H^l_{kl}M^i_m ( \mathcal{B}_8^A)^{kmj} + B^A_{21}(\mathcal{B}_{c6})_{ij} H^l_{kl}M^k_m ( \mathcal{B}_8^A)^{imj}\nonumber\\
& + B^A_{22}(\mathcal{B}_{c6})_{ij} H^l_{kl}M^m_m ( \mathcal{B}_8^A)^{ikj} + B^A_{23}(\mathcal{B}_{c6})_{ij} H^l_{lk}M^i_m ( \mathcal{B}_8^A)^{jkm}+ B^A_{24}(\mathcal{B}_{c6})_{ij} H^l_{lk}M^i_m ( \mathcal{B}_8^A)^{jmk}\nonumber\\
& + B^A_{25}(\mathcal{B}_{c6})_{ij} H^l_{lk}M^i_m ( \mathcal{B}_8^A)^{kmj} + B^A_{26}(\mathcal{B}_{c6})_{ij} H^l_{lk}M^k_m ( \mathcal{B}_8^A)^{imj}+ B^A_{27}(\mathcal{B}_{c6})_{ij} H^l_{lk}M^m_m ( \mathcal{B}_8^A)^{ikj}.
\end{align}
The total amplitude for the $\mathcal{B}_{c6}\to \mathcal{B}_8 M$ decays is
\begin{align}
  \mathcal{A}(\mathcal{B}_{c6}\to \mathcal{B}_8 M) = \mathcal{A}^S(\mathcal{B}_{c6}\to \mathcal{B}_8^S M) + \mathcal{A}^A(\mathcal{B}_{c6}\to \mathcal{B}_8^A M).
\end{align}
The topological diagrams contributing to $\mathcal{B}_{c6}\to \mathcal{B}_8^S M$ decays are shown in Fig.~\ref{top7}.
The corresponding diagrams for $\mathcal{B}_{c6}\to \mathcal{B}_8^A M$ decays can be obtained by replacing the symmetric quark pairs with antisymmetric ones in diagrams $B_1^S\sim B_{27}^S$.
The topological amplitudes for the $\Omega^0_c\to \mathcal{B}_{8}^SM$ and $\Omega^0_c\to \mathcal{B}_{8}^AM$ decays are presented in Tables~\ref{amp15} and \ref{amp16}, respectively.

The octet baryon can also be represented as a tensor with one covariant index and one contravariant index, $(\mathcal{B}_8)^i_j$, where $i\neq j$.
It can also be expressed in matrix form as
\begin{eqnarray}
 \mathcal{B}_8=  \left( \begin{array}{ccc}
   \frac{1}{\sqrt 2} \Sigma^0+  \frac{1}{\sqrt 6} \Lambda^0    & \Sigma^+  & p \\
    \Sigma^- &   - \frac{1}{\sqrt 2} \Sigma^0+ \frac{1}{\sqrt 6} \Lambda^0   & n \\
    \Xi^- & \Xi^0 & -\sqrt{2/3}\Lambda^0 \\
  \end{array}\right).
\end{eqnarray}
The third-rank tensors $(\mathcal{B}^S_8)^{ijk}$ and $(\mathcal{B}^A_8)^{ijk}$ can be expressed in terms of the rank-$(1,1)$ tensor as \cite{Wang:2025bdl,Wang:2024ztg}
\begin{eqnarray}\label{sy1}
(\mathcal{B}^S_8)^{ijk} = \frac{1}{\sqrt{6}}\left[\epsilon^{kil}(\mathcal{B}_8)^j_l
+\epsilon^{kjl}(\mathcal{B}_8)^i_l\right],\qquad\quad  (\mathcal{B}^A_8)^{ijk}= \frac{1}{\sqrt{2}}\epsilon^{ijl}(\mathcal{B}_8)^k_l.
\end{eqnarray}
The indices $i$ and $j$ are symmetric under the interchange $i \leftrightarrow j$ in $(\mathcal{B}^S_8)^{ijk}$ and antisymmetric in $(\mathcal{B}^A_8)^{ijk}$.
To study the linear relations between the decay amplitudes for $\mathcal{B}_{c6}\to \mathcal{B}_{8}^SM$ and $\mathcal{B}_{c6}\to \mathcal{B}_{8}^AM$, we construct the decay amplitude for $\mathcal{B}_{c6}\to \mathcal{B}_{8}M$ using the rank-$(1,1)$ octet tensor as
\begin{align}\label{amp4}
 \mathcal{A}(\mathcal{B}_{c6}\to \mathcal{B}_8M) &=  B_1(\mathcal{B}_{c6})_{ij} H^m_{kl}M^i_m ( \mathcal{B}_8)^{l}_p\epsilon^{jkp} + B_2(\mathcal{B}_{c6})_{ij} H^m_{kl}M^i_m ( \mathcal{B}_8)^{k}_p\epsilon^{jlp}+ B_3(\mathcal{B}_{c6})_{ij}H^m_{kl}M^i_m ( \mathcal{B}_8)^{j}_p\epsilon^{klp} \nonumber\\
 & + B_4(\mathcal{B}_{c6})_{ij} H^i_{kl}M^j_m (\mathcal{B}_8)^{m}_p\epsilon^{klp}   + B_5(\mathcal{B}_{c6})_{ij} H^i_{kl}M^j_m (\mathcal{B}_8)^{l}_p\epsilon^{kmp}   + B_6(\mathcal{B}_{c6})_{ij} H^i_{kl}M^j_m ( \mathcal{B}_8)^{k}_p\epsilon^{lmp}  \nonumber\\
 &+ B_7(\mathcal{B}_{c6})_{ij} H^i_{kl}M^k_m (\mathcal{B}_8)^{m}_p\epsilon^{jlp} + B_8(\mathcal{B}_{c6})_{ij} H^i_{kl}M^k_m (\mathcal{B}_8)^{l}_p\epsilon^{jmp}  + B_9(\mathcal{B}_{c6})_{ij} H^i_{kl}M^k_m ( \mathcal{B}_8)^{j}_p\epsilon^{lmp}  \nonumber\\
 &+ B_{10}(\mathcal{B}_{c6})_{ij} H^i_{kl}M^l_m ( \mathcal{B}_8)^{m}_p\epsilon^{jkp}   + B_{11}(\mathcal{B}_{c6})_{ij} H^i_{kl}M^l_m (\mathcal{B}_8)^{k}_p\epsilon^{jmp}  + B_{12}(\mathcal{B}_{c6})_{ij} H^i_{kl}M^l_m ( \mathcal{B}_8)^{j}_p\epsilon^{kmp}  \nonumber\\
&+ B_{13}(\mathcal{B}_{c6})_{ij} H^m_{kl}M^l_m ( \mathcal{B}_8)^{j}_p\epsilon^{ikp} + B_{14}(\mathcal{B}_{c6})_{ij} H^m_{kl}M^k_m (\mathcal{B}_8)^{j}_p\epsilon^{ilp}  + B_{15}(\mathcal{B}_{c6})_{ij} H^i_{kl}M^m_m ( \mathcal{B}_8)^{l}_p\epsilon^{jkp}  \nonumber\\
 &+ B_{16}(\mathcal{B}_{c6})_{ij} H^i_{kl}M^m_m ( \mathcal{B}_8)^{k}_p\epsilon^{jlp}  + B_{17}(\mathcal{B}_{c6})_{ij} H^i_{kl}M^m_m (\mathcal{B}_8)^{j}_p\epsilon^{klp}   + B_{18}(\mathcal{B}_{c6})_{ij} H^l_{kl}M^i_m ( \mathcal{B}_8)^{m}_p\epsilon^{jkp} \nonumber\\
 & + B_{19}(\mathcal{B}_{c6})_{ij} H^l_{kl}M^i_m ( \mathcal{B}_8)^{k}_p\epsilon^{jmp}  + B_{20}(\mathcal{B}_{c6})_{ij} H^l_{kl}M^i_m ( \mathcal{B}_8)^{j}_p\epsilon^{kmp}  + B_{21}(\mathcal{B}_{c6})_{ij} H^l_{kl}M^k_m ( \mathcal{B}_8)^{j}_p\epsilon^{imp} \nonumber\\
& + B_{22}(\mathcal{B}_{c6})_{ij} H^l_{kl}M^m_m ( \mathcal{B}_8)^{j}_p\epsilon^{ikp}  + B_{23}(\mathcal{B}_{c6})_{ij} H^l_{lk}M^i_m ( \mathcal{B}_8)^{m}_p\epsilon^{jkp} + B_{24}(\mathcal{B}_{c6})_{ij} H^l_{lk}M^i_m ( \mathcal{B}_8)^{k}_p\epsilon^{jmp} \nonumber\\
& + B_{25}(\mathcal{B}_{c6})_{ij} H^l_{lk}M^i_m ( \mathcal{B}_8)^{j}_p\epsilon^{kmp}  + B_{26}(\mathcal{B}_{c6})_{ij} H^l_{lk}M^k_m ( \mathcal{B}_8)^{j}_p\epsilon^{imp} + B_{27}(\mathcal{B}_{c6})_{ij} H^l_{lk}M^m_m ( \mathcal{B}_8)^{j}_p\epsilon^{ikp} .
\end{align}
The linear relations among the amplitudes constructed from the rank-3 and (1,1)-rank octet tensors are
\begin{align}\label{sol2}
  B_1 & = B^A_1-B^S_2+B^S_3,\qquad B_2 = B^A_2-B^S_1+B^S_3,\qquad B_3 = B^A_3-B^S_1+B^S_2,\qquad B_4 = B^A_4-B^S_5+B^S_6,\nonumber\\
  B_5 & = B^A_5-B^S_4+B^S_6,\qquad B_6 = B^A_6-B^S_4+B^S_5,\qquad B_7 = B^A_7-B^S_8+B^S_9,\qquad B_8 = B^A_8-B^S_7+B^S_9,\nonumber\\
  B_9 & = B^A_9-B^S_7+B^S_8,\qquad B_{10} = B^A_{10}-B^S_{11}+B^S_{12},\qquad B_{11} = B^A_{11}-B^S_{10}+B^S_{12},\qquad B_{12} = B^A_{12}-B^S_{10}+B^S_{11},\nonumber\\
  B_{13} & = B^A_{13}+B^S_{13}-2B^S_{29},\qquad B_{14} = B^A_{14}+B^S_{14}-2B^S_{28},\qquad B_{15} = B^A_{15}-B^S_{16}+B^S_{17},\qquad B_{16} = B^A_{16}-B^S_{15}+B^S_{17},\nonumber\\
  B_{17} & = B^A_{17}-B^S_{15}+B^S_{16},\qquad B_{18} = B^A_{18}-B^S_{19}+B^S_{20},\qquad B_{19} = B^A_{19}-B^S_{18}+B^S_{20},\qquad B_{20} = B^A_{20}-B^S_{18}+B^S_{19},\nonumber\\
  B_{21} & = B^A_{21}+B^S_{21}-2B^S_{30},\qquad B_{22} = B^A_{22}+B^S_{22}-2B^S_{31},\qquad B_{23} = B^A_{23}-B^S_{24}+B^S_{25},\qquad B_{24} = B^A_{24}-B^S_{23}+B^S_{25},\nonumber\\
  B_{25} & = B^A_{25}-B^S_{23}+B^S_{24},\qquad B_{26} = B^A_{26}+B^S_{26}-2B^S_{32},\qquad B_{27} = B^A_{27}+B^S_{27}-2B^S_{33}.
\end{align}

The $SU(3)$ irreducible amplitude for the $\mathcal{B}_{c6}\to \mathcal{B}_{8} M$ decay can be expressed as
\begin{align}\label{amp5}
 \mathcal{A}^{IR}(\mathcal{B}_{c6}\to \mathcal{B}_8M) &=  b^{15}_1(\mathcal{B}_{c6})_{ij} H(15)^m_{kl}M^i_m ( \mathcal{B}_8)^{l}_p\epsilon^{jkp} + b^{\overline 6}_1(\mathcal{B}_{c6})_{ij} H(\overline6)^m_{kl}M^i_m ( \mathcal{B}_8)^{l}_p\epsilon^{jkp} \nonumber\\
 &+ b^{\overline 6}_2(\mathcal{B}_{c6})_{ij}H(\overline6)^m_{kl}M^i_m ( \mathcal{B}_8)^{j}_p\epsilon^{klp}+ b^{\overline 6}_3(\mathcal{B}_{c6})_{ij} H(\overline6)^i_{kl}M^j_m (\mathcal{B}_8)^{m}_p\epsilon^{klp} \nonumber\\
 & + b^{15}_2(\mathcal{B}_{c6})_{ij} H(15)^i_{kl}M^j_m (\mathcal{B}_8)^{l}_p\epsilon^{kmp}  + b^{\overline 6}_4(\mathcal{B}_{c6})_{ij} H(\overline6)^i_{kl}M^j_m (\mathcal{B}_8)^{l}_p\epsilon^{kmp} \nonumber\\
 &+ b^{15}_3(\mathcal{B}_{c6})_{ij} H(15)^i_{kl}M^k_m (\mathcal{B}_8)^{m}_p\epsilon^{jlp}+ b^{15}_4(\mathcal{B}_{c6})_{ij} H(15)^i_{kl}M^k_m (\mathcal{B}_8)^{l}_p\epsilon^{jmp} \nonumber\\
 & + b^{15}_5(\mathcal{B}_{c6})_{ij} H(15)^i_{kl}M^k_m ( \mathcal{B}_8)^{j}_p\epsilon^{lmp}+ b^{\overline 6}_{5}(\mathcal{B}_{c6})_{ij} H(\overline6)^i_{kl}M^k_m (\mathcal{B}_8)^{m}_p\epsilon^{jlp}
  \nonumber\\
 &+ b^{\overline 6}_{6}(\mathcal{B}_{c6})_{ij} H(\overline6)^i_{kl}M^k_m (\mathcal{B}_8)^{l}_p\epsilon^{jmp}
  + b^{\overline 6}_{7}(\mathcal{B}_{c6})_{ij} H(\overline6)^i_{kl}M^k_m ( \mathcal{B}_8)^{j}_p\epsilon^{lmp} \nonumber\\
&+ b^{15}_{6}(\mathcal{B}_{c6})_{ij} H(15)^m_{kl}M^l_m ( \mathcal{B}_8)^{j}_p\epsilon^{ikp}+ b^{\overline 6}_{8}(\mathcal{B}_{c6})_{ij} H(\overline6)^m_{kl}M^l_m ( \mathcal{B}_8)^{j}_p\epsilon^{ikp}  \nonumber\\
 &+ b^{15}_{7}(\mathcal{B}_{c6})_{ij} H(15)^i_{kl}M^m_m ( \mathcal{B}_8)^{l}_p\epsilon^{jkp} + b^{\overline 6}_{9}(\mathcal{B}_{c6})_{ij} H(\overline6)^i_{kl}M^m_m ( \mathcal{B}_8)^{l}_p\epsilon^{jkp} \nonumber\\
 &+ b^{\overline 6}_{10}(\mathcal{B}_{c6})_{ij} H(\overline6)^i_{kl}M^m_m (\mathcal{B}_8)^{j}_p\epsilon^{klp}  + b^3_{1}(\mathcal{B}_{c6})_{ij} H(3)_{k}M^i_m ( \mathcal{B}_8)^{m}_p\epsilon^{jkp}\nonumber\\
 & + b^3_{2}(\mathcal{B}_{c6})_{ij} H(3)_{k}M^i_m ( \mathcal{B}_8)^{k}_p\epsilon^{jmp} + b^3_{3}(\mathcal{B}_{c6})_{ij} H(3)_{k}M^i_m ( \mathcal{B}_8)^{j}_p\epsilon^{kmp}\nonumber\\
 & + b^3_{4}(\mathcal{B}_{c6})_{ij} H(3)_{k}M^k_m ( \mathcal{B}_8)^{j}_p\epsilon^{imp} + b^3_{5}(\mathcal{B}_{c6})_{ij} H(3)_{k}M^m_m ( \mathcal{B}_8)^{j}_p\epsilon^{ikp}
\nonumber\\
&+ b^{3^\prime}_{1}(\mathcal{B}_{c6})_{ij} H(3^\prime)_{k}M^i_m ( \mathcal{B}_8)^{m}_p\epsilon^{jkp}+ b^{3^\prime}_{2}(\mathcal{B}_{c6})_{ij} H(3^\prime)_{k}M^i_m ( \mathcal{B}_8)^{k}_p\epsilon^{jmp}\nonumber\\
 &+ b^{3^\prime}_{3}(\mathcal{B}_{c6})_{ij} H(3^\prime)_{k}M^i_m ( \mathcal{B}_8)^{j}_p\epsilon^{kmp}+ b^{3^\prime}_{4}(\mathcal{B}_{c6})_{ij} H(3^\prime)_{k}M^k_m ( \mathcal{B}_8)^{j}_p\epsilon^{imp}\nonumber\\
& + b^{3^\prime}_{5}(\mathcal{B}_{c6})_{ij} H(3^\prime)_{k}M^m_m ( \mathcal{B}_8)^{j}_p\epsilon^{ikp} .
\end{align}
The linear relations between the $SU(3)$ irreducible amplitudes and the rank-($1,1$) amplitudes are
\begin{align}\label{sol3}
   & b^{15}_1 = B_1+B_2,\quad b^{\overline 6}_1 = B_1-B_2,  \quad b^{\overline 6}_2 = B_3,  \quad b^{\overline 6}_3 = B_4, \quad b^{15}_2 = B_5+B_6, \quad  b^{\overline 6}_4 = B_5-B_6,\nonumber\\
   & b^{15}_3 = B_7+B_{10},\quad b^{\overline 6}_{5} = B_7-B_{10},  \quad b^{15}_4 = B_8 + B_{11},  \quad b^{\overline 6}_{6} = B_8 - B_{11}, \quad b^{15}_5 = B_9+B_{12}, \quad  b^{\overline 6}_{7} = B_9-B_{12},\nonumber\\
   & b^{15}_{6} = B_{13}+B_{14},\quad b^{\overline 6}_{8} = B_{13}-B_{14},  \quad b^{15}_{7} = B_{15} + B_{16},  \quad b^{\overline 6}_{9} = B_{15} - B_{16}, \quad b^{\overline 6}_{10} = B_{17},\nonumber\\
   &  b^3_{1} = \frac{3}{8}B_1- \frac{1}{8}B_2 - \frac{1}{2}B_4- \frac{1}{8}B_7 +\frac{3}{8}B_{10}+ B_{18},\quad b^{3^\prime}_{1} = -\frac{1}{8}B_1+ \frac{3}{8}B_2 + \frac{1}{2}B_4+ \frac{3}{8}B_7 -\frac{1}{8}B_{10}+ B_{23},\nonumber\\
   &  b^3_{2} = -\frac{1}{8}B_1+ \frac{3}{8}B_2 - \frac{1}{8}B_{5}+ \frac{3}{8}B_{6} -\frac{1}{8}B_{8}+ \frac{3}{8}B_{11}+ B_{19},\nonumber\\& b^{3^\prime}_{2} = \frac{3}{8}B_1- \frac{1}{8}B_2 + \frac{3}{8}B_{5}- \frac{1}{8}B_{6} +\frac{3}{8}B_{8}- \frac{1}{8}B_{11}+ B_{24},\nonumber\\
   &  b^3_{3} = \frac{1}{2}B_3+ \frac{3}{8}B_5 - \frac{1}{8}B_{6}- \frac{1}{8}B_{9} +\frac{3}{8}B_{12}+ B_{20},\quad b^{3^\prime}_{3} = -\frac{1}{2}B_3- \frac{1}{8}B_5 + \frac{3}{8}B_{6}+ \frac{3}{8}B_{9} -\frac{1}{8}B_{12}+ B_{25},
\nonumber\\
   &  b^3_{4} = \frac{3}{8}B_8- \frac{1}{8}B_{11} - \frac{1}{8}B_{12} -\frac{1}{8}B_{13} + \frac{3}{8}B_{14}+B_{21},\nonumber\\& b^{3^\prime}_{4} = -\frac{1}{8}B_8+ \frac{3}{8}B_{11} +\frac{3}{8}B_{12} +\frac{3}{8}B_{13} - \frac{1}{8}B_{14}+B_{26},
\nonumber\\
   &  b^3_{5} = \frac{3}{8}B_{13}- \frac{1}{8}B_{14} +\frac{3}{8}B_{15}- \frac{1}{8}B_{16} +\frac{1}{2}B_{17} +B_{22},\quad  b^{3^\prime}_{5} = -\frac{1}{8}B_{13}+\frac{3}{8}B_{14} -\frac{1}{8}B_{15}+\frac{3}{8}B_{16} -\frac{1}{2}B_{17} +B_{27}.
\end{align}
Using the Levi-Civita symbol, the ten amplitudes constructed from the $\overline 6$ representation in Eq.~\eqref{amp5} are simplified to five terms as
\begin{align}\label{amp6}
 \mathcal{A}^{IR}(\mathcal{B}_{c6}\to \mathcal{B}_8M) &=  b^{\prime\overline 6}_{1}(\mathcal{B}_{c6})_{ij} H(\overline 6)^{ij}M^k_l ( \mathcal{B}_8)^{l}_k + b^{\prime\overline 6}_2(\mathcal{B}_{c6})_{ij} H(\overline6)^{kl}M^i_k ( \mathcal{B}_8)^{j}_l + b^{\prime\overline 6}_3(\mathcal{B}_{c6})_{ij}H(\overline6)^{ik}M^j_l ( \mathcal{B}_8)^{l}_k\nonumber\\
 &+ b^{\prime\overline 6}_4(\mathcal{B}_{c6})_{ij} H(\overline6)^{ik}M^l_k (\mathcal{B}_8)^{j}_l+ b^{\prime\overline 6}_5(\mathcal{B}_{c6})_{ij} H(\overline6)^{ik}M^l_l (\mathcal{B}_8)^{j}_k.
\end{align}
The relations among the amplitudes constructed from the $\overline 6$ representation are
\begin{align}
 b^{\prime\overline 6}_{1} & = -b^{\overline 6}_5-b^{\overline 6}_6,\qquad  b^{\prime\overline 6}_{2} = -b^{\overline 6}_1+2b^{\overline 6}_2-b^{\overline 6}_8,\qquad b^{\prime\overline 6}_{3} = 2b^{\overline 6}_3+b^{\overline 6}_4+b^{\overline 6}_5+b^{\overline 6}_6,\nonumber\\
 b^{\prime\overline 6}_{4} & = b^{\overline 6}_6+b^{\overline 6}_7+b^{\overline 6}_8,\qquad  b^{\prime\overline 6}_{5} = -b^{\overline 6}_6-b^{\overline 6}_7-b^{\overline 6}_9+2b^{\overline 6}_{10}.
\end{align}
The $SU(3)$ irreducible amplitudes involving $H^{(0)}(3)$, $H^{(1)}(3)$, and $H^{(1)}(3^\prime)$ appear only in the following fixed combinations:
\begin{align}\label{qx1}
  b_{1}^{3,T+P} &= b^3_{1} + b^{3P}_{1} + 3b^{3^\prime P}_{1},\qquad
   b_{2}^{3,T+P} = b^3_{2} + b^{3P}_{2} + 3b^{3^\prime P}_{2}, \qquad
   b_{3}^{3,T+P} = b^3_{3} + b^{3P}_{3} + 3b^{3^\prime P}_{3},\nonumber\\
  b_{4}^{3,T+P} &= b^3_{4} + b^{3P}_{4} + 3b^{3^\prime P}_{4},\qquad
  b_{5}^{3,T+P} = b^3_{5} + b^{3P}_{5} + 3b^{3^\prime P}_{5}.
\end{align}
Substituting Eq.~\eqref{sol3} into Eq.~\eqref{qx1}, we find that the penguin-induced diagrams $B_{1}^P\sim B_{27}^P$ appear together with the tree-induced diagrams $B_{18}\sim B_{22}$ in the following fixed combinations:
\begin{align}
  B_{18}^{T+P} &= B_{18} + B^P_2 + B^P_4+ B^P_7 +  B^P_{18} +3B^P_{23},\qquad B_{19}^{T+P}= B_{19} + B^P_1 + B^P_5+ B^P_8 +  B^P_{19} +3B^P_{24}, \nonumber\\
   B_{20}^{T+P} &= B_{20} - B^P_3 + B^P_6+ B^P_9 +  B^P_{20} +3B^P_{25},\qquad B_{21}^{T+P}= B_{21} + B^P_{11} + B^P_{12}+ B^P_{13} +  B^P_{21} +3B^P_{26}, \nonumber\\
B_{22}^{T+P} &= B_{22} + B^P_{14} + B^P_{16}- B^P_{17} +  B^P_{22}
+3B^P_{27}.
\end{align}
In total, 20 independent amplitudes contribute to $\mathcal{B}_{c6}\to \mathcal{B}_{8}M$ decays.

\section{Phenomenological analysis}\label{pa}

The $\Omega^0_c$ is an isospin singlet.
We identify three isospin sum rules for $\Omega^0_c$ decays:
\begin{align}
  \mathcal{A}(\Omega^0_c\to \Delta^+K^-)- \mathcal{A}(\Omega^0_c\to \Delta^0\overline K^0)=0,
\end{align}
\begin{align}
  \mathcal{A}(\Omega^0_c\to \Sigma^{*+}\pi^-)+2\mathcal{A}(\Omega^0_c\to \Sigma^{*0}\pi^0)-\mathcal{A}(\Omega^0_c\to \Sigma^{*-}\pi^+)=0,
\end{align}
\begin{align}
  \mathcal{A}(\Omega^0_c\to \Sigma^{+}\pi^-)-2\mathcal{A}(\Omega^0_c\to \Sigma^{0}\pi^0)+\mathcal{A}(\Omega^0_c\to \Sigma^{-}\pi^+)=0.
\end{align}
$\Omega^0_c$ is a member of the $U$-spin triplet $(\Sigma_c^0,\Xi_c^{*0},\Omega^0_c)$.
Because the two neutral baryons in the same $U$-spin multiplet as $\Omega^0_c$ can undergo strong decays, $U$-spin sum rules cannot be established for this system.

Seven tree-level diagrams contribute to the $\mathcal{B}_{c6}\to \mathcal{B}_{10} M$ decays.
Diagram $A_7$ is suppressed because it requires the exchange of two gluons.
For diagrams $A_1\sim A_6$, the corresponding Wilson coefficients are given by
\begin{align}
 A_1:\,\,\,C_2(\mu)+C_1(\mu)/3,\qquad   A_2:\,\,\,C_1(\mu)+C_2(\mu)/3, \qquad   A_3:\,\,\,(C_1(\mu)+C_2(\mu))/3,\nonumber\\
 A_4:\,\,\,(C_2(\mu)+C_1(\mu))/3,\qquad   A_5:\,\,\,C_2(\mu)+C_1(\mu)/3, \qquad   A_6:\,\,\,C_1(\mu)+C_2(\mu)/3.
\end{align}
The Wilson coefficients $C_1(\mu)$ and $C_2(\mu)$ at $\mu=m_c$ are given in the NDR scheme by $C_1(m_c)\approx-0.43$ and $C_2(m_c)\approx1.22$ \cite{Li:2012cfa}.
This yields
\begin{align}
  C_2(\mu)+C_1(\mu)/3\approx 1.08,\qquad C_1(\mu)+C_2(\mu)/3\approx-0.02,\qquad (C_1(\mu)+C_2(\mu))/3\approx 0.26.
\end{align}
Thus, diagrams $A_2$, $A_3$, $A_4$, and $A_6$ are suppressed by the corresponding Wilson coefficients.
The branching fractions for decay modes involving $A_1$ and $A_5$ are expected to be larger than those for the other decay modes.
Neglecting the contribution from diagram $A_6$, the ratio $\mathcal{B}r(\Omega^0_c\to \Omega^-K^+)/\mathcal{B}r(\Omega^0_c\to \Omega^-\pi^+)$ can be estimated as
\begin{align}\label{r}
  \frac{\mathcal{B}r(\Omega^0_c\to \Omega^-K^+)}{\mathcal{B}r(\Omega^0_c\to \Omega^-\pi^+)}\simeq\frac{|V_{us}|^2}{|V_{ud}|^2}\approx 5.33\times 10^{-2}.
\end{align}
The LHCb Collaboration reported \cite{LHCb:2023fvd} that
\begin{align}
  \frac{\mathcal{B}r(\Omega^0_c\to \Omega^-K^+)}{\mathcal{B}r(\Omega^0_c\to \Omega^-\pi^+)}=(6.08\pm 0.51\pm0.40)\times 10^{-2},
\end{align}
This is consistent with Eq.~\eqref{r}.
Moreover, the ratio $\mathcal{B}r(\Omega^0_c\to \Xi^{*-}\pi^+)/\mathcal{B}r(\Omega^0_c\to \Omega^-\pi^+)$ is predicted to be
\begin{align}
  \frac{\mathcal{B}r(\Omega^0_c\to \Xi^{*-}\pi^+)}{\mathcal{B}r(\Omega^0_c\to \Omega^-\pi^+)}\simeq\frac{1}{3}\frac{|V_{us}|^2}{|V_{ud}|^2}\approx 1.78\times 10^{-2}.
\end{align}
From an experimental perspective, the ratio of the branching fractions of the two Cabibbo-favored decay modes, $\mathcal{B}r(\Omega^0_c\to \Xi^0\overline K^0)/\mathcal{B}r(\Omega^0_c\to \Omega^-\pi^+)$, has been measured by the Belle Collaboration \cite{Belle:2017szm}.
We suggest measuring a further ratio of branching fractions for two Cabibbo-favored decay modes, $\mathcal{B}r(\Omega^0_c\to \Xi^{*0}\overline K^0)/\mathcal{B}r(\Omega^0_c\to \Omega^-\pi^+)$, where the $\Xi^{*0}$ baryon is reconstructed through the decay chain $\Xi^{*0}\to \Xi^-[\to \Lambda^0(\to p \pi^-)\pi^-]\pi^+$.
Since the decay $\Omega^0_c\to \Sigma^{*+} K^-$ is dominated by diagram $A_5$, we also suggest measuring the ratio $\mathcal{B}r(\Omega^0_c\to \Sigma^{*+} K^-)/\mathcal{B}r(\Omega^0_c\to \Omega^-\pi^+)$, where the $\Sigma^{*+}$ baryon is reconstructed through $\Sigma^{*+}\to \Lambda^0(\to p \pi^-)\pi^+$.

\begin{figure}
  \centering
  \includegraphics[width=14cm]{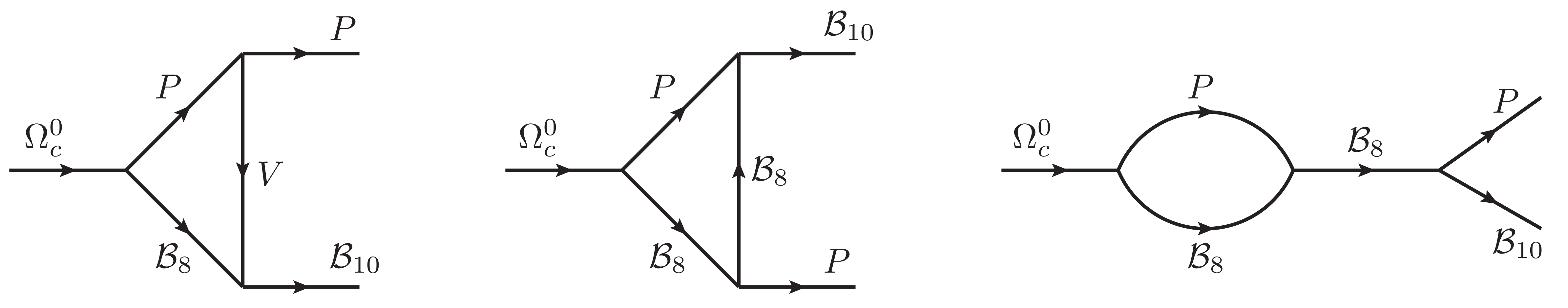}
  \caption{The $u$-, $t$-, and $s$-channel rescattering contributions to $\Omega^0_c\to\mathcal{B}_{10}M$ decays }\label{top1}
\end{figure}

Owing to the limited experimental data currently available, the magnitudes and strong phases of the topological amplitudes cannot yet be determined.
To gain further insight into these amplitudes, we examine their relationship with the rescattering mechanism.
In this mechanism, a charmed baryon first decays into a baryon and a meson through a short-distance emission diagram.
Subsequently, the $u$-, $t$-, and $s$-channel meson-baryon scattering processes provide the long-distance contributions.
The $u$-, $t$-, and $s$-channel rescattering contributions to the $\Omega^0_c\to\mathcal{B}_{10}M$ decays are shown in Fig.~\ref{top1}.
The theoretical framework for studying the relationship between topological diagrams and the rescattering mechanism has been developed in Refs.~\cite{Wang:2022wrb,Lai:2025ryi}.
The leading-order chiral Lagrangians for the $VPP$, $\mathcal{B}_8\mathcal{B}_8P$, $\mathcal{B}_{10}\mathcal{B}_8P$, and $\mathcal{B}_{10}\mathcal{B}_8V$ vertices are given by \cite{Scherer:2002tk,Meissner:1987ge,Bernard:1995dp,Aliev:2010aj,Holmberg:2018dtv}
\begin{align}\label{x6}
\mathcal{L}_{\mathcal{V}\mathcal{P}\mathcal{P}} &= \frac{ig_{VPP}}{\sqrt{2}}
\,Tr\,(\mathcal{V}^{\mu}\,[\mathcal{P},\partial_\mu \mathcal{P}]),\qquad
  \mathcal{L}_{\mathcal{B}\mathcal{B}\mathcal{P}}= \sqrt{2}D \,Tr(\overline {\mathcal{B}}\,i\gamma_5\{ \mathcal{P},\mathcal{B}\}) + \sqrt{2}F \,Tr(\overline {\mathcal{B}}\,i\gamma_5[ \mathcal{P},\mathcal{B}]),\nonumber\\
   \mathcal{L}_{\mathcal{T}\mathcal{B}\mathcal{P}}&=  g_{TBP}\epsilon_{ijk}\overline {\mathcal{T}}^{kml}_{\mu}(\partial^{\mu}\mathcal{P})^i_m\mathcal{B}^{j}_l,\qquad
   \mathcal{L}_{\mathcal{T}\mathcal{B}\mathcal{V}}=   g_{TBV}\epsilon_{ijk}\overline   {\mathcal{T}}^{kml}_{\mu}(\mathcal{V}^\mu)^i_m\mathcal{B}^{j}_l,
\end{align}
where $\mathcal{P}$, $\mathcal{V}$, $\mathcal{B}$, and $\mathcal{T}$ denote the matrices for the pseudoscalar mesons, vector mesons, octet baryons, and decuplet baryons, respectively.
In the tensor formalism, the amplitudes for the $VPP$, $\mathcal{B}_8\mathcal{B}_8P$, $\mathcal{B}_{10}\mathcal{B}_8P$, and $\mathcal{B}_{10}\mathcal{B}_8V$ strong interactions can be expressed as
\begin{align}\label{x7}
  \mathcal{A}_{VPP}& = \alpha^+P^j_i V^i_k P^k_j +\alpha^-P^j_i V^k_jP^i_k,\qquad \mathcal{A}_{\mathcal{B}_8\mathcal{B}_8P} = \beta^+(\mathcal{B}_8)^j_i P^i_k (\mathcal{B}_8)^k_j +\beta^-(\mathcal{B}_8)^j_i P^k_j (\mathcal{B}_8)^i_k,\nonumber\\
 \mathcal{A}_{\mathcal{B}_{10}\mathcal{B}_8P} &= \zeta \, (\mathcal{B}_{10})^{kml}P^i_m(\mathcal{B}_8)^{j}_l \epsilon_{ijk},\qquad
 \mathcal{A}_{\mathcal{B}_{10}\mathcal{B}_8V} = \kappa \, (\mathcal{B}_{10})^{kml}V^i_m(\mathcal{B}_8)^{j}_l \epsilon_{ijk}.
\end{align}
Comparison of Eq.~\eqref{x6} with Eq.~\eqref{x7} shows that the coupling coefficients $\alpha^{\pm}$, $\beta^{\pm}$, $\zeta$, and $\kappa$ correspond to those in the chiral Lagrangians as follows:
\begin{align}
  \alpha^+&=-\alpha^-=\frac{ig_{VPP}}{\sqrt{2}}V^\mu P \partial_\mu P,\qquad
  \beta^+=\sqrt{2}(D+F)\,\overline {\mathcal{B}}_8\,i\gamma_5  P\mathcal{B}_8,\qquad \beta^-=\sqrt{2}(D-F)\,\overline {\mathcal{B}}_8\,i\gamma_5 \mathcal{B}_8 P,\nonumber\\
 \zeta&=g_{TBP}(\overline{\mathcal{B}}_{10})_\mu(\partial^{\mu}P)\mathcal{B}_8,\qquad
  \kappa=g_{TBV}(\overline{\mathcal{B}}_{10})_\mu V^\mu\mathcal{B}_8.
\end{align}

\begin{table*}[t!]
\caption{Rescattering amplitudes contributing to $\mathcal{B}_{c6}\to \mathcal{B}_{10}M$ decays. }\label{tab}
\begin{tabular}{|c|c|c|c|}
\hline\hline
 & ~~~~$u$-channel~~~~  & ~~~~$t$-channel~~~~ & ~~~~$s$-channel~~~~  \\\hline
 $~~A_2~~$ &  $\Delta_{\alpha^+,\kappa}$  &  $-\Delta_{\beta^-,\zeta}$  &    \\\hline
 $A_3$ &    & $-\Delta_{\beta^+,\zeta}$   &  $-\Theta_{\beta^+,\zeta}$  \\\hline
 $A_4$ &  $-\Delta_{\alpha^+,\kappa}$  &   $\Delta_{\beta^-,\zeta}$  &    \\\hline
 $A_5$ &   &  $\Delta_{\beta^-,\zeta}$  &  $\Theta_{(\beta^+-\beta^-),\zeta}$  \\\hline
 $A_6$ &  $-\Delta_{\alpha^-,\kappa}$  &   & $\Theta_{\beta^-,\zeta}$   \\\hline
 $A_8$ &   &  $\Delta_{(\beta^+-\beta^-),\zeta}$  &  $\Theta_{\beta^-,\zeta}$  \\\hline
 $A_9$ &  $\Delta_{\alpha^-,\kappa}$  & &  $-\Theta_{\beta^-,\zeta}$  \\\hline
\hline
\end{tabular}
\end{table*}

For the $\mathcal{B}_{c6}\to \mathcal{B}_8M$ decays, the color-supported emission diagram is denoted by $B_{28}^S$.
According to Eq.~\eqref{sol2}, $B_{28}^S$ contributes only to the amplitude $B_{14}$.
Thus, the amplitude $B_{14}$ serves as the weak vertex for the dominant rescattering contributions.
In the $SU(3)_F$ limit, the rescattering amplitudes can be obtained through tensor contractions.
For the $u$-channel rescattering process $\mathcal{B}_{c6}\to P^\prime\mathcal{B}^\prime_8\to P\mathcal{B}_{10}$, with a vector meson serving as the intermediate propagator, the tensor structure is given by
\begin{align}
 \mathcal{R}^{u}_{10} &=\sum_{P^\prime,\mathcal{B}^\prime_8,V}(\mathcal{B}_{c6})_{ij}H^m_{kl}
 (P^\prime)^k_m(\mathcal{B}^\prime_8)^j_n\epsilon^{iln}\times[\alpha^+(P^\prime)^b_aV^a_cP^c_b
  +\alpha^-(P^\prime)^a_bV^c_aP^b_c] \times \kappa \, (\mathcal{B}_{10})^{qpf}V^d_p(\mathcal{B}^\prime_8)^{e}_q \epsilon_{def}.
\end{align}
The tensor structure for the $t$-channel rescattering process $\mathcal{B}_{c6}\to P^\prime\mathcal{B}^\prime_8\to P\mathcal{B}_{10}$, mediated by an octet baryon $\mathcal{B}^{\prime\prime}_8$, is given by
\begin{align}
  \mathcal{R}^{t}_{10}&=  \sum_{P^\prime,\mathcal{B}^\prime,\mathcal{B}^{\prime\prime}}(\mathcal{B}_{c6})_{ij}H^m_{kl}
 (P^\prime)^k_m(\mathcal{B}^\prime_8)^j_n\epsilon^{iln}\times
 [\beta^+(\mathcal{B}^{\prime})^a_c(\mathcal{B}^{\prime\prime})^b_aP^c_b
  +\beta^-(\mathcal{B}^{\prime})^c_a(\mathcal{B}^{\prime\prime})^a_bP^b_c]
 \times \zeta \, (\mathcal{B}_{10})^{qpf}(P^\prime)^d_p(\mathcal{B}^{\prime\prime}_8)^{e}_q \epsilon_{def}.
\end{align}
The tensor structure for the $s$-channel rescattering process $\mathcal{B}_{c6}\to P^\prime\mathcal{B}^\prime_8\to P\mathcal{B}_{10}$ mediated by an octet baryon $\mathcal{B}^{\prime\prime}_8$ is given by
\begin{align}
 \mathcal{R}^{s}_{10}&= \sum_{P^\prime,\mathcal{B}^\prime,\mathcal{B}^{\prime\prime}}(\mathcal{B}_{c6})_{ij}H^m_{kl}
 (P^\prime)^k_m(\mathcal{B}^\prime_8)^j_n\epsilon^{iln}
 \times[\beta^+(P^\prime)^a_b(\mathcal{B}^\prime)^c_a
 (\mathcal{B}^{\prime\prime})^b_c
  +\beta^-(P^\prime)^b_a(\mathcal{B}^\prime)^a_c(\mathcal{B}^{\prime\prime})^c_b]
  \times\zeta \, (\mathcal{B}_{10})^{qpf}P^d_p(\mathcal{B}^{\prime\prime}_8)^{e}_q \epsilon_{def}.
\end{align}
With the completeness relations
\begin{align}
  \sum_{P^\prime}(P^\prime)^i_j(P^\prime)^k_l & = \delta^i_l\delta^k_j-\frac{1}{3}\delta^i_j\delta^k_l,\qquad \sum_{\mathcal{B}^{\prime(\prime)}_{8}}(\mathcal{B}^{\prime(\prime)}_{8})^i_j
  (\mathcal{B}^{\prime(\prime)}_{8})^k_l = \delta^i_l\delta^k_j-\frac{1}{3}\delta^i_j\delta^k_l,\qquad
 \sum_{V}V^i_jV^k_l = \delta^i_l\delta^k_j,
\end{align}
We derive the rescattering amplitudes that contribute to the topological amplitudes of the $\mathcal{B}_{c6}\to \mathcal{B}_{10}M$ decays in the $SU(3)_F$ limit.
The results are presented in Table~\ref{tab}.
We use "$\Delta$" to denote a triangle diagram and "$\Theta$" to denote a bubble diagram.
The subscripts of $\Delta$ and $\Theta$ denote the two strong vertices of the corresponding triangle and bubble diagrams.
Note that the triangle and bubble diagrams include contributions from both ground-state and excited hadrons.

According to Table~\ref{tab} and the theoretical calculations of triangle diagrams \cite{Yu:2017zst,Han:2021azw,Hu:2024uia,Jia:2024pyb,Hu:2026drh}, the quark-loop diagrams receive long-distance contributions of the same order of magnitude as the tree diagrams, whereas the color-favored emission diagram appears to be about one order of magnitude larger than the other diagrams.
Thus, if a singly Cabibbo-suppressed channel does not involve the color-favored emission diagram, the CP asymmetry in this channel is expected, at the naive level, to be of order $\mathcal{O}(10^{-4}\sim 10^{-3})$.
If the color-favored emission diagram contributes to a singly Cabibbo-suppressed channel, the CP asymmetry in this channel would be suppressed.
In the future, once sufficient measurements of branching fractions and decay parameters become available, the nonperturbative parameters of the tree amplitudes can be extracted from experimental data.
Using the relationship between the topological diagrams and the final-state rescattering mechanism, these parameters can then be used to estimate the quark-loop amplitudes and, consequently, the corresponding CP asymmetries.

In the literature, the K\"orner-Pati-Woo theorem \cite{Pati:1970fg,Korner:1970xq} is often used to simplify the decay amplitudes of heavy baryon decays.
According to the K\"orner-Pati-Woo theorem, if the two quarks produced by the weak operators enter the same baryon, they must be antisymmetric in flavor.
For $\Omega_c^0\to \mathcal{B}_{10}M$ decays, the two quark lines emitted from the weak vertex enter the final-state baryon in diagrams $A_3$, $A_4$, $A_5$, $A_6$, and $A_7$.
Therefore, these quarks must be antisymmetric in flavor according to the K\"orner-Pati-Woo theorem.
The $\mathcal{O}(\overline{6})$ and $\mathcal{O}(15)$ operators are antisymmetric and symmetric, respectively, under the interchange of two covariant flavor indices.
Thus, only the $\mathcal{O}(\overline 6)$ operators contribute to the diagrams $A_3$, $A_4$, $A_5$, $A_6$, and $A_7$.
According to Eq.~\eqref{sol}, we have
\begin{align}
 A_3 = A_4 = A_7 =0,\qquad A_5 = -A_6.
\end{align}
The above equations yield the following relations in addition to the general isospin sum rules:
 \begin{align}\label{q1}
 \mathcal{B}r(\Omega_{c}^{0}\to \Sigma^{*+}K^-) = 4\,\mathcal{B}r(\Omega_{c}^{0}\to \Sigma^{*0}K^0_S),
\end{align}
 \begin{align}
 \mathcal{B}r(\Omega_{c}^{0}\to \Sigma^{*+}\pi^-) = \mathcal{B}r(\Omega_{c}^{0}\to \Sigma^{*0}\pi^0)=\mathcal{B}r(\Omega_{c}^{0}\to \Sigma^{*-}\pi^+),
\end{align}
 \begin{align}\label{q3}
 \mathcal{B}r(\Omega_{c}^{0}\to \Delta^{+}K^-) = \mathcal{B}r(\Omega_{c}^{0}\to \Delta^{0}K^0_S)=0.
\end{align}
For the decays $\Omega_c^0\to \mathcal{B}_{8}M$, the K\"orner-Pati-Woo theorem yields the following relations among the topological diagrams:
\begin{align}
 B_1^{S,A} &= -B_2^{S,A}, \qquad  B_5^{S,A} = -B_6^{S,A}, \qquad B_7^{S,A} = -B_{10}^{S,A}, \qquad  B_8^{S,A} = -B_{11}^{S,A},\nonumber\\
 B_9^{S,A} &= -B_{12}^{S,A}, \qquad  B_{15}^{S,A} = -B_{16}^{S,A}, \qquad B_3^{S} = B_{4}^{S}=B_{17}^{S}=0.
\end{align}
These lead to the following relations, which go beyond the general isospin sum rules:
 \begin{align}\label{q2}
 \mathcal{B}r(\Omega_{c}^{0}\to \Sigma^{+}K^-) = 4\,\mathcal{B}r(\Omega_{c}^{0}\to \Sigma^{0}K^0_S),
\end{align}
 \begin{align}
 \mathcal{B}r(\Omega_{c}^{0}\to \Sigma^{+}\pi^-) = \mathcal{B}r(\Omega_{c}^{0}\to \Sigma^{0}\pi^0)=\mathcal{B}r(\Omega_{c}^{0}\to \Sigma^{-}\pi^+),
\end{align}
 \begin{align}
 \mathcal{B}r(\Omega_{c}^{0}\to pK^-) = 2\,\mathcal{B}r(\Omega_{c}^{0}\to nK^0_S),
\end{align}
 \begin{align}\label{q4}
 \mathcal{B}r(\Omega_{c}^{0}\to \Sigma^{0}\eta^{(\prime)})=\mathcal{B}r(\Omega_{c}^{0}\to \Lambda^{0}\pi^0)=0.
\end{align}
Ref.~\cite{Lai:2025ryi} showed that the K\"orner-Pati-Woo theorem is inconsistent with the rescattering dynamics of singly charmed baryon decays.
The present work further supports this conclusion.
For example, the K\"orner-Pati-Woo theorem predicts $A_3=A_4=0$, which conflicts with the nonvanishing contributions of the triangle and bubble diagrams to $A_3$ and $A_4$.
In the isospin limit, the K\"orner-Pati-Woo theorem can be tested experimentally using Eqs.~\eqref{q1}$\sim$ \eqref{q3} and Eqs.~\eqref{q2}$\sim$ \eqref{q4}.
If the decay amplitudes of two channels are proportional to each other, their decay parameters are identical.
Thus, the K\"orner-Pati-Woo theorem can also be tested by measuring the decay parameters $\alpha$, $\beta$, and $\gamma$.
In addition, if the K\"orner-Pati-Woo theorem holds, two CP-asymmetry relations can be derived in the isospin limit:
\begin{align}
    A_{CP}^{\rm dir}(\Omega^0_c\to \Sigma^{*+}K^-)=A_{CP}^{\rm dir}(\Omega^0_c\to \Sigma^{*0}\overline K^0),\qquad A_{CP}^{\rm dir}(\Omega^0_c\to \Sigma^{+}K^-)=A_{CP}^{\rm dir}(\Omega^0_c\to \Sigma^{0}\overline K^0).
\end{align}
The partial-wave CP asymmetries can also be defined as
\begin{align}
  A_{CP}^{\alpha} = \frac{\alpha+\overline\alpha}{2}, \qquad A_{CP}^{\beta} = \frac{\beta+\overline\beta}{2},\qquad A_{CP}^{\gamma} = \frac{\gamma-\overline\gamma}{2}.
\end{align}
The following relations can be derived under isospin symmetry if the K\"orner-Pati-Woo theorem holds:
\begin{align}
    A_{CP}^{\alpha}(\Omega^0_c\to \Sigma^{*+}K^-)&=A_{CP}^{\alpha}(\Omega^0_c\to \Sigma^{*0}\overline K^0),\qquad A_{CP}^{\alpha}(\Omega^0_c\to \Sigma^{+}K^-)=A_{CP}^{\alpha}(\Omega^0_c\to \Sigma^{0}\overline K^0),\\
  A_{CP}^{\beta}(\Omega^0_c\to \Sigma^{*+}K^-)&=A_{CP}^{\beta}(\Omega^0_c\to \Sigma^{*0}\overline K^0),\qquad A_{CP}^{\beta}(\Omega^0_c\to \Sigma^{+}K^-)=A_{CP}^{\beta}(\Omega^0_c\to \Sigma^{0}\overline K^0),\\
    A_{CP}^{\gamma}(\Omega^0_c\to \Sigma^{*+}K^-)&=A_{CP}^{\gamma}(\Omega^0_c\to \Sigma^{*0}\overline K^0),\qquad A_{CP}^{\gamma}(\Omega^0_c\to \Sigma^{+}K^-)=A_{CP}^{\gamma}(\Omega^0_c\to \Sigma^{0}\overline K^0).
\end{align}
These CP-asymmetry relations can be tested in future measurements.

\section{Summary}\label{summary}

In this work, we analyze the topological amplitudes for $\Omega_c^0$ decays in the $SU(3)_F$ limit.
We present the tree- and penguin-induced diagrams contributing to $\Omega_c^0$ decays into decuplet and octet baryons.
Using tensor analysis, we derive the linear relations between the topological amplitudes and the $SU(3)$ irreducible amplitudes.
We obtain three isospin relations for $\Omega^0_c$ decays.
In addition, we derive several branching-fraction relations to test the K\"orner-Pati-Woo theorem.

\begin{acknowledgements}

This work was supported in part by Scientific Research Fund of Hunan Provincial Department under No. 25B0090.

\end{acknowledgements}

\end{document}